# Origins of extreme emission-line ratios in $z > 3$ galaxies: insights from the LUMEN model

Lucie Scharré ,[1★] Michaela Hirschmann ,[1,2] Adèle Plat,[1] Stephane Charlot ,[3] Rachel S. Somerville ,[4] Emma Curtis-Lake ,[5] Gabriella De Lucia ,[2] Miroslava Dessauges-Zavadsky ,[6] Anna Feltre ,[7] Marion Farcy ,[1] Pascale Jablonka,[1,8] Natalia Lahén ,[9,10] Aswin P. Vijayan[11] and Stephen M. Wilkins [11]

[1]*Institute of Physics, Laboratory for Galaxy Evolution, École Polytechnique Fédérale de Lausanne (EPFL), Observatoire de Sauverny, CH-1290 Versoix, Switzerland*
[2]*INAF – Osservatorio Astronomico di Trieste, Via G. B. Tiepolo 11, I-34143 Trieste, Italy*
[3]*Sorbonne Université, CNRS, UMR 7095, Institut d'Astrophysique de Paris, 98 bis bd Arago, F-75014 Paris, France*
[4]*Center for Computational Astrophysics, Flatiron Institute, New York, NY 10010, USA*
[5]*Centre for Astrophysics Research, Department of Physics, Astronomy and Mathematics, University of Hertfordshire, Hatfield AL10 9AB, UK*
[6]*Département d'Astronomie, Université de Genève, Chemin Pegasi 51, CH-1290 Versoix, Switzerland*
[7]*INAF – Osservatorio Astrofisico di Arcetri, Largo E. Fermi 5, I-50125 Florence, Italy*
[8]*LIRA, Observatoire de Paris, Université PSL, Sorbonne Université, Université Paris Cité, CY Cergy Paris Université, CNRS, F-92190 Meudon, France*
[9]*Zentrum für Astronomie der Universität Heidelberg, Astronomisches Rechen-Institut, Mönchhofstr. 12-14, D-69120 Heidelberg, Germany*
[10]*Max-Planck-Institut für Astrophysik, Karl-Schwarzschild-Str. 1, D-85748, Garching, Germany*
[11]*Astronomy Centre, University of Sussex, Falmer, Brighton BN1 9QH, UK*

Accepted 2026 September 1. Received 2026 August 1; in original form 2026 May 7

## ABSTRACT

Optical emission-line ratios in star-forming galaxies at $z \sim 3$–8, such as [O III]/H $\beta$ and [O III]/[O II], are strongly offset from those at $z \sim 0$–2, pointing to more extreme ionization and ISM conditions in the early Universe. To constrain the physical origin of these offsets, we developed LUMEN, a framework for modelling nebular emission from spatially distributed H II regions in cosmological simulations. We apply LUMEN to ILLUSTRISTNG50, validate its predictions at low redshift, and test mechanisms proposed to explain extreme line ratios at $z = 3$–8. We focus on the [N II]/H $\alpha$ versus [O III]/H $\beta$ (N2-BPT) diagram, the [S II]/H $\alpha$ versus [O III]/H $\beta$ (S2-VO87) diagram, and the [O III]/[O II] versus ([O II]+[O III])/H $\beta$ (O32–R23) diagram. We find that stellar $\alpha$-enhancement alone cannot explain the bulk of observations. Moderate offsets arise from the combined effects of stellar $\alpha$-enhancement and a higher IMF upper-mass cutoff, while AGNs may contribute to some of the observed scatter. The most extreme [O III]/H $\beta$ and [O III]/[O II] values require high ionization parameters powered by massive star clusters of $\gtrsim 10^5$–$10^6$ M$_\odot$, consistent with recent *JWST* observations. Reproducing the highest [N II]/H $\alpha$ ratios requires enhanced nitrogen abundances. Although gas densities of $n \sim 10^4$ cm$^{-3}$ can boost several diagnostic ratios, they suppress [S II]/H $\alpha$, in tension with current observations. Overall, a model combining harder ionizing spectra, elevated ionization parameters from massive star clusters, and enhanced nitrogen abundances can reproduce all observed high-$z$ line ratios across the N2-BPT, S2-VO87, and O32–R23 diagrams. Based on this, we provide new demarcation lines for star-forming galaxies in the N2-BPT and S2-VO87 diagrams.

## 1 INTRODUCTION

Nebular emission lines arising from the ionized gas in the interstellar medium (ISM) of galaxies are a product of both the local gas conditions and the nature of the ionizing sources. Gas density, chemical abundances, and the shape and intensity of the ionizing spectrum are encoded in the relative strengths of hydrogen recombination lines and collisionally excited metal lines (G. J. Ferland & H. Netzer 1983; D. E. Osterbrock & G. J. Ferland 2006, see L. J. Kewley, D. C. Nicholls & R. S. Sutherland 2019 for a review). Diagnostic diagrams built from ratios of rest-frame optical lines, such as [N II]λ6584, H $\alpha$, [O I]λ6300, [O III]λ5007, H $\beta$, and the doublets [O II]λλ3727, 3729 and [S II]λλ6717, 6731, can therefore serve as powerful tools to constrain galaxy properties.

The standard Baldwin–Phillips–Terlevich (BPT; J. A. Baldwin, M. M. Phillips & R. Terlevich 1981) and Veilleux–Osterbrock (VO87; S. Veilleux & D. E. Osterbrock 1987) diagrams connect the [O III]$\lambda$5007/H$\beta$ (R3) ratio to the [N II]$\lambda$6584/H$\alpha$ (N2), [S II]$\lambda$6724/H$\alpha$ (S2), and [O I]$\lambda$6300/H$\alpha$ ratios in order to identify the dominant ionizing sources in nearby galaxies. In these diagrams, local H II regions and star-forming galaxies are confined to distinct regions, visually separated from signatures that are dominated by line emission originating from active galactic nuclei (AGNs; L. J. Kewley et al. 2001a; G. Kauffmann et al. 2003).

★ E-mail: lucie.scharre@epfl.ch

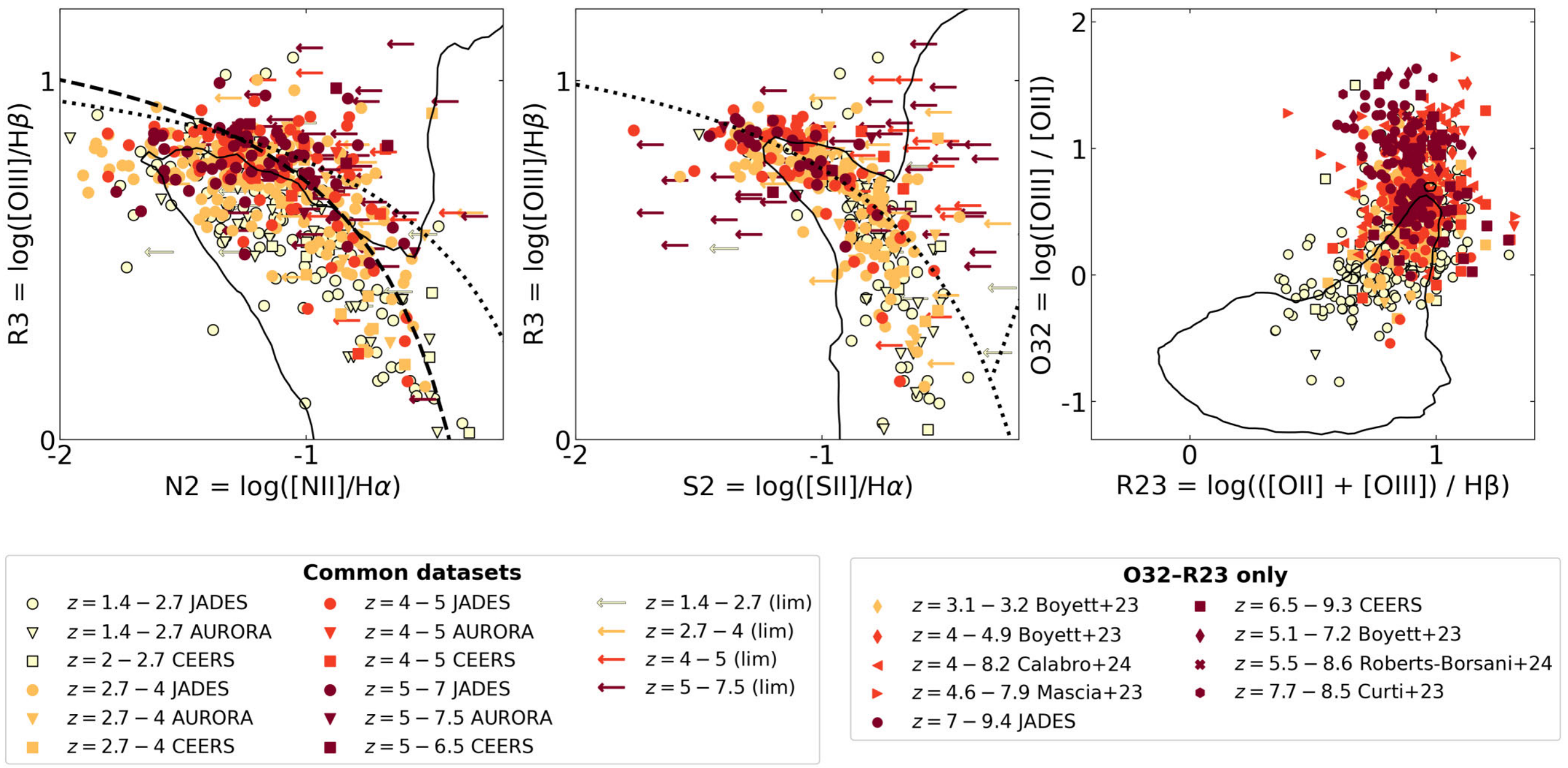

**Figure 1.** Compilation of star-forming galaxy observations between $z = 1.4$ and $z = 9.4$ in the N2-BPT (left), S2-VO87 (middle), and O32–R23 (right) diagrams, as indicated in the legend. Included for comparison are local SDSS-observed galaxies (black outline), the theoretical upper limits for star-forming galaxies (L. J. Kewley et al. 2001a; dotted lines), and the empirical lower limit for AGNs (G. Kauffmann et al. 2003; dashed line).

For the N2-BPT diagram, it is well established that line ratios from star-forming galaxies observed at $z \sim 1$–2 are offset from the sequence occupied by their local counterparts (e.g. X. Liu et al. 2008; F. Bian et al. 2010; D. Masters et al. 2014; C. C. Steidel et al. 2014; A. E. Shapley et al. 2015; D. Kashino et al. 2017; T. Kojima et al. 2017; A. L. Strom et al. 2017; J. N. Runco et al. 2021). Similarly, in the S2-VO87 diagram, galaxies observed at $z \sim 2$ are generally shifted towards higher S2 at fixed R3 compared to local H II regions (A. E. Shapley et al. 2019; F. Mannucci et al. 2021). Given the diagonal sequence of the star-forming galaxies across both BPT diagrams, the observed offset can be interpreted as a shift towards higher R3 at fixed N2 or S2, or the reverse.

The [O III]$\lambda$5007/[O II]$\lambda\lambda$3727, 3729 (O32) versus ([O II]$\lambda\lambda$3727, 3729 + [O III]$\lambda\lambda$4959, 5007)/H$\beta$ (R23) diagram is another popular diagnostic tool for which pre-*JWST* observations have indicated an evolution with redshift. Locally, it correlates the ionization state, traced by O32, with the gas-phase metallicity, traced by R23. At $z \sim 2$, both O32 and R23 values appear elevated compared to those produced by nearby galaxies (e.g. A. E. Shapley et al. 2015, 2019; R. L. Sanders et al. 2016; C. C. Steidel et al. 2016; A. L. Strom et al. 2017, 2018; M. W. Topping et al. 2020; J. N. Runco et al. 2021). Together, these offsets indicate that the physical conditions of galaxies at cosmic noon differ significantly enough to systematically alter the relative strengths of optical emission lines.

The *James Webb Space Telescope* (*JWST*) recently extended measurements of rest-frame optical emission lines beyond the ground-based near-IR cut-off at $z \sim 2.7$, thanks to new high-quality spectra collected by the NIRSpec instrument (A. J. Cameron et al. 2023, 2026; S. Mascia et al. 2023; K. Nakajima et al. 2023; F. Sun et al. 2023; R. L. Sanders et al. 2023a; K. Boyett et al. 2024; G. Roberts-Borsani et al. 2024; A. Calabrò et al. 2024a, b; K. Z. Arellano-Córdova et al. 2025; A. E. Shapley et al. 2025; M. W. Topping et al. 2025b; L. Clarke et al. 2026). This has enabled direct tests of whether emission-line properties continue evolving at earlier epochs. Fig. 1 shows a comprehensive compilation of *JWST* observations in the N2-BPT (left panel), S2-VO87 (middle panel), and O32–R23 (right panel) diagrams. Redshift increases from light yellow at $z \lesssim 2.7$ (outlined in black) to dark red out to $z = 9.4$. Arrows indicate upper limits. For reference, we show local data from the Sloan Digital Sky Survey (SDSS; black outline).

Fig. 1 demonstrates that star-forming galaxies at $z > 2.7$ in the N2-BPT diagram are even further offset from the local population than the $z \leq 2.7$ galaxies, indicating a continuing evolution. This was one of the main results from the Assembly of Ultradeep Rest-optical Observations Revealing Astrophysics (AURORA; triangular points; A. E. Shapley et al. 2025) survey, which presented a sample of 52 emission-line galaxies in the N2-BPT diagram between $z = 1.4$ and 7.5. The *JWST* Advanced Deep Extragalactic Survey (JADES; circular points) confirmed this trend using a sample of over 300 individual galaxies, in which galaxies at $z \gtrsim 3.5$ are offset relative to those at $z \sim 2$ (F. D'Eugenio et al. 2025; A. J. Cameron et al. 2026; L. Clarke et al. 2026). We further include earlier data from JADES (A. J. Cameron et al. 2023) and the Cosmic Evolution Early Release Science (CEERS; square data points; R. L. Sanders et al. 2026) survey, which similarly contain extreme R3 and N2 values out to $z = 6.5$.

[S II]$\lambda$6724 detections in the AURORA, JADES, and CEERS surveys have populated the high-$z$ S2-VO87 diagram as well. Above $z = 2.7$, the data are shifted towards higher R3 and S2 relative to lower redshifts. In the O32–R23 plane, the evolution

of line ratios also persists beyond cosmic noon, where galaxies with increasing redshift produce progressively higher O32 values (see legend; A. J. Cameron et al. 2023; S. Mascia et al. 2023; K. Nakajima et al. 2023; R. L. Sanders et al. 2023a; K. Boyett et al. 2024; G. Roberts-Borsani et al. 2024; A. Calabrò et al. 2024a, b; L. Clarke et al. 2026).

These new *JWST* observations reveal that the local gas and ionization conditions of galaxies beyond cosmic noon are likely even more extreme than their $z \lesssim 2.7$ counterparts. Several physical processes have been proposed to explain both the $z \sim 2$ offset from the local population and the more extreme line ratios now observed at higher redshift. For example, generally higher star formation rates (SFRs) and densities may increase the ionization parameter and thereby boost high-ionization lines like [O III]$\lambda$5007 in O32 and R3 (J. Brinchmann, M. Pettini & S. Charlot 2008; L. J. Kewley et al. 2013a, b; C. C. Steidel et al. 2014; M. Hirschmann et al. 2017; D. Kashino et al. 2017; T. Kojima et al. 2017; A. L. Strom et al. 2017; M. Hirschmann et al. 2019; D. Kashino et al. 2019; A. J. Cameron et al. 2023; M. Hirschmann et al. 2023b; Euclid Collaboration 2024). Harder ionizing spectra from binary stars, a higher upper-mass cutoff of the initial mass function (IMF), $\alpha$-element-enhanced stellar populations or unidentified AGN contributions, could boost the high-energy photon budget and lead to higher excitation in all line ratios (X. Liu et al. 2008; S. A. Wright et al. 2010; C. C. Steidel et al. 2016; M. W. Topping et al. 2020; J. N. Runco et al. 2021; R. L. Sanders et al. 2023a; A. E. Shapley et al. 2025). Variations in chemical abundances, particularly elevated N/O ratios, could enhance N2 at fixed metallicity (D. Masters et al. 2014; A. E. Shapley et al. 2015; R. L. Sanders et al. 2016). Lyman continuum (LyC) leaking density-bounded H II regions with a non-zero $f_{\rm esc}$ are expected to decrease low-ionization lines with respect to high-ionization lines, which could boost O32 (A. E. Jaskot & M. S. Oey 2013; K. Nakajima & M. Ouchi 2014; A. Plat et al. 2019; Y. I. Izotov et al. 2021; S. R. Flury et al. 2022a, b; A. J. Cameron et al. 2023; S. Mascia et al. 2023; R. L. Sanders et al. 2023b; K. Boyett et al. 2024; A. Calabrò et al. 2024a, b). To constrain which mechanism might be responsible for the observed line signatures at high redshift, it is crucial to understand each of these effects and their potential degeneracies in detail.

Theoretical frameworks for predicting nebular emission in galaxy formation models offer controlled settings to test competing physical scenarios. One straightforward approach is to make use of simulations that include radiative transfer modelling. Based on its local ionization state, the gas can be post-processed with photoionization models, as has been done for the SPHINX simulations (H. Katz et al. 2022b, 2023; N. Choustikov et al. 2024) and the THESAN-ZOOM suite (W. McClymont, A. Smith & S. Tacchella 2025). The MEGATRON suite and its predecessors (H. Katz 2022; H. Katz et al. 2022a, 2025; N. Choustikov et al. 2026; H. Katz et al. 2026) even combine their radiative transfer with a nebular emission prediction entirely on the fly. While these models couple the ionizing radiation field self-consistently to the gas, they require high resolution and full radiation-hydrodynamics, and are therefore limited in cosmological volume and redshift coverage.

Modelling nebular emission without explicitly including radiative transfer allows for larger cosmological volumes and improved statistics. Semi-analytic models and most cosmological simulations generally do not resolve the multiphase ISM from which emission lines originate. As a result, studies often predict the integrated galaxy emission by relying on galaxy-wide properties to assign each galaxy one photoionization model meant to represent the contributions from all H II regions. Such methods have been applied to semi-analytic models such as SAGS, GALACTICUS, L-GALAXIES, GALFORM, and GAEA (A. Orsi et al. 2014; V. Gonzalez-Perez et al. 2018; A. Merson et al. 2018; D. Izquierdo-Villalba et al. 2019; Z. Zhai et al. 2019; G. Favole et al. 2020; V. Gonzalez-Perez et al. 2020; C. M. Baugh et al. 2022; A. Knebe et al. 2022; Euclid Collaboration 2024), as well as hydrodynamical simulations such as ILLUSTRISTNG (M. Hirschmann, S. Charlot & R. S. Somerville 2023a; M. Hirschmann et al. 2023b). While these models enable large statistical samples, they ignore any internal variation of the emission-line properties. Yet observations from *JWST*, ALMA, and integral field unit spectroscopy show that the ionized ISM is highly inhomogeneous at high redshift, with strong line-ratio variations driven by local differences in density, metallicity, and ionization state (X. Ji et al. 2024; M. Usui et al. 2025; D. A. Berg et al. 2026).

Some of this internal diversity is captured by methods that model H II regions based on the properties of individual star particles. This strategy has been applied to BLUETIDES, ILLUSTRISTNG, TECHNICOLORDAWN, FLARES, FIRSTLIGHT, and CIELO (I. Shimizu et al. 2016; X. Shen et al. 2020; S. M. Wilkins et al. 2020; D. Ceverino et al. 2021; A. P. Vijayan et al. 2021; Y. Nakazato, N. Yoshida & D. Ceverino 2023; S. M. Wilkins et al. 2023; A. Cornejo-Cárdenas et al. 2025). However, star particles are typically assigned a fixed mass ($\gtrsim 10^5$ $M_\odot$), whereas observed star clusters exhibit a wide range of masses. As a result, individual H II-region properties are not well represented in these methods.

To bridge this resolution gap, the WARPFIELD framework (E. W. Pellegrini et al. 2020) explicitly samples the observed star-cluster mass function down to 100 $M_\odot$ and places clusters within local density peaks. Applied to a Milky Way-like zoom simulation from the AURIGA project at $z = 0$ (R. J. J. Grand et al. 2016), this approach yields spatially resolved H II-region emission. A similar strategy is adopted by P. Garg et al. (2022) for the SIMBA simulations at $z = 0$ and 2, where star particles are subdivided into lower-mass clusters. However, in their method, the emission-line predictions are based solely on the properties of the parent star particle, and thus the star clusters are not coupled to their local gas environment.

Among these existing frameworks, none is able to (i) retain the cosmological context and statistical power of large-volume simulations from low to high redshift, (ii) capture the diversity of nebular conditions within galaxies, and (iii) produce realistic H II-region properties. We address this gap with LUMEN, our emission-line framework which places star clusters in cosmological simulations to enable consistent predictions for both individual H II regions and integrated galaxy populations at any redshift. Inspired by E. W. Pellegrini et al. (2020), LUMEN resamples the young star particles in each galaxy into star clusters drawn from the cluster mass function found in the high-resolution GRIFFIN simulation (N. Lahén et al. 2020), which is consistent with observational determinations. After placing these star clusters into the dense gas of the original galaxy, each is assigned a stellar population and CLOUDY (G. J. Ferland et al. 2017) photoionization model. In applying LUMEN to the ILLUSTRISTNG50 simulation (A. Pillepich et al. 2018), we obtained large statistical samples of emission-line galaxies in a cosmological context. This provides an ideal testing ground for the impact of different physical mechanisms and gas conditions on nebular emission, thereby constraining the origin of observed elevated line ratios.

This paper is structured as follows. In Section 2, we review the physical mechanisms thought to influence optical emission-line ratios. Section 3 describes the LUMEN methodology and the suite of model variations designed to represent these mechanisms. We validate our Base model against key observables at $z = 0$ in Section 4. In Section 5, we assess whether the model variations can reproduce the line ratios observed at $z \geq 2.7$. We discuss implications and limitations of our study in Section 6, and summarize our conclusions in Section 7. The notation used throughout this paper is provided in Table 1, where each abbreviation is defined as the logarithm of the line ratio.

**Table 1.** Definition of line ratios adopted throughout this paper.

| Notation | Line ratio |
|---|---|
| R3 | log ([O III] λ5007/Hβ) |
| N2 | log ([N II] λ6584/Hα) |
| S2 | log ([S II] λλ6717, 6731/Hα) |
| R2 | log ([O II] λλ3727, 3729/Hβ) |
| R23 | log (([O II] λλ3727, 3729 + [O III] λλ4959, 5007)/Hβ) |
| O32 | log ([O III] λ5007/[O II] λλ3727, 3729) |
| N2O2 | log ([N II] λ6584/[O II] λλ3727, 3729) |

## 2 MECHANISMS THAT COULD ALTER LINE RATIOS

We first describe the physical background of different ISM and ionizing source conditions commonly inferred to explain variation in emission-line properties. Each is modelled in a separate LUMEN version (see Section 3.4). We assess their impact on our emission-line predictions in Section 5.

### 2.1 Increased hardness of the ionizing spectrum

The ionizing spectrum may be hardened by variations in stellar population properties or the presence of AGNs. The resulting increase in high-energy photons can elevate R3, N2, S2, and O32. Harder ionizing spectra from $\alpha$-enhanced stellar populations are a widely discussed explanation for the observed line-ratio offset at $z \sim 2$ and beyond (C. C. Steidel et al. 2016; A. L. Strom et al. 2017; A. E. Shapley et al. 2019, 2025; R. L. Sanders et al. 2020, 2023a; M. W. Topping et al. 2020; J. N. Runco et al. 2021). Direct constraints from rest-frame UV stellar features have shown that massive stellar populations in galaxies at cosmic noon can indeed exhibit elevated $\alpha$/Fe ratios (F. Cullen et al. 2021; D. Kashino et al. 2022; A. L. Strom et al. 2022; T. M. Stanton et al. 2024). Such abundance patterns arise because core-collapse supernovae enrich the ISM rapidly with $\alpha$ elements, while a substantial fraction of Fe enrichment from Type Ia supernovae occurs on longer delay time-scales (C. Kobayashi, A. I. Karakas & M. Lugaro 2020). This effect is particularly relevant at high redshift, where galaxies are observed closer to the onset of star formation and therefore may not yet have experienced sufficient delayed Type Ia enrichment to lower their $\alpha$/Fe ratios towards local values. At fixed total metallicity, higher stellar $\alpha$/Fe ratios reduce the opacity of stellar atmospheres, which produces hotter, more compact stars and increases the output of high-energy UV photons. Additional mechanisms such as binary star interactions, like mass transfer and mergers, can extend the stellar lifetime and create hot helium stars, which also harden the stellar spectral energy distribution (SED). This effect is likely enhanced at high redshift, as younger, metal-poor stellar populations contain more massive stars (J. J. Eldridge & E. R. Stanway 2012; J. J. Eldridge et al. 2017; Y. Götberg et al. 2020). Similarly, the presence of more massive stars due to a higher IMF upper-mass cutoff may also harden the stellar radiation. The impact of both has been discussed in C. C. Steidel et al. (2016) and A. Plat et al. (2019). Lastly, AGNs generally produce harder radiation than stellar populations. While a strong AGN component can usually be identified and excluded, a weak undetected contribution may also harden the overall ionizing spectrum and affect the observed line emission (X. Liu et al. 2008; S. A. Wright et al. 2010). We include an implementation of each of these effects among our suite of LUMEN models.

### 2.2 Increased ionization parameters

Higher ionization parameters at high redshift have also been proposed to explain elevated R3 and O32 ratios (J. Brinchmann et al. 2008; L. J. Kewley et al. 2013a, b; C. C. Steidel et al. 2014; M. Hirschmann et al. 2017; D. Kashino et al. 2017; T. Kojima et al. 2017; A. L. Strom et al. 2017; M. Hirschmann et al. 2019; D. Kashino et al. 2019; A. J. Cameron et al. 2023; N. A. Reddy et al. 2023; M. Hirschmann et al. 2023b; Euclid Collaboration 2024). The ionization parameter quantifies the strength of an ionizing radiation field relative to the local gas density. In spherical geometry, the ionization parameter $U$ at the Strömgren radius $R_{\rm S}$ (itself given by equation 1) can be defined as in equation (2) (see e.g. S. Charlot & M. Longhetti 2001).

$$R_{\rm S}^3 = \frac{3Q}{4\pi n_{\rm H}^2 \epsilon \alpha_{\rm B}} \tag{1}$$

$$U = \frac{\alpha_{\rm B}^{2/3}}{3c}\left(\frac{3Q\epsilon^2 n_{\rm H}}{4\pi}\right)^{1/3}. \tag{2}$$

Here, an ionizing photon rate $Q$ illuminates dense hydrogen clumps with density $n_{\rm H}$. The volume-filling factor $\epsilon$ specifies the fraction of the nebular volume occupied by these clumps. $\alpha_{\rm B}$ is the case-B hydrogen recombination coefficient (D. E. Osterbrock & G. J. Ferland 2006). Equation (2) shows that an increase in the ionization parameter may be linked to an increase in the ionizing photon flux $Q$, the ionized gas density $n_{\rm H}$, or the volume-filling factor $\epsilon$. Each of these is discussed below.

#### *2.2.1 Increasing the ionizing photon flux Q via increased star-cluster masses*

An increase in $Q$ represents a greater number of ionizing photons emitted per unit time. The value of $Q$ depends primarily on the age and mass of the star cluster, with younger and more massive clusters emitting significantly more photons than older and less massive populations. We propose that the higher ionization parameters inferred at high redshift may be linked to H II regions powered by systematically more massive stellar clusters. Observations increasingly indicate that high-$z$ star formation is compact and clumpy, with stellar clumps showing median masses of order $10^7$ M$_\odot$ between $z = 0.7$ and 5.5 (M. Dessauges-Zavadsky et al. 2017; A. Claeyssens et al. 2023; M. Messa et al. 2024; A. Claeyssens et al. 2025). While the sample of resolved high-$z$ star clusters remains limited, these systems may be more massive than typical young star clusters in nearby galaxies. In the local Universe, star-cluster masses commonly span $\sim 10^3$–$10^5$ M$_\odot$, with only a subset reaching above $10^5$ M$_\odot$ (M. R. Krumholz, C. F. McKee & J. Bland-Hawthorn 2019; D. Pathak et al. 2025). By contrast, *JWST* observations of strongly lensed galaxies have revealed compact star clusters and cluster candidates with masses of $\sim 10^5$–$10^6$ M$_\odot$ at $z \gtrsim 2$, with some detected out to $z \sim 10$ (A. Claeyssens et al. 2023; E. Vanzella et al. 2023; A. Adamo et al. 2024; L. Mowla et al. 2024; Abdurro'uf et al. 2025; A. Claeyssens et al. 2025, 2026; M. Messa et al. 2026).

In the first statistical analysis of high-$z$ star clusters, A. Claeyssens et al. (2026) find a cluster mass-function slope compatible with the canonical $\beta \simeq -2$ form above the estimated completeness limit of $\sim 1.5 \times 10^6$ $M_\odot$. It remains unclear, however, how many lower-mass clusters remain undetected because of resolution and sensitivity limits. The Cosmic Gems galaxy, a compact galaxy at $z = 9.625$ containing five clusters with masses $> 10^6$ $M_\odot$, provides a useful constraint (A. Adamo et al. 2024). E. Vanzella et al. (2026) argue that if the star cluster population followed a local-like mass function extending to masses as low as $\sim 10^2$–$10^3$ $M_\odot$, the implied total cluster mass would exceed the stellar mass budget of the host. This system may therefore exhibit a higher low-mass cutoff in the star-cluster mass function or a flatter mass-function slope. Similar arguments have been made for other strongly lensed high-$z$ systems, suggesting that this may not be unique to Cosmic Gems (Messa et al., in preparation).

The interpretation that star clusters at high redshift are systematically more massive is also supported by theoretical models, which predict that the high gas surface densities, turbulence, and pressures characteristic of early galaxies favour more massive bound clusters (J. M. Kruijssen 2015; M. Reina-Campos & J. M. Kruijssen 2017; G. De Lucia et al. 2024a, and references therein). These massive star clusters would naturally increase $Q$ and $U$ per H II region, potentially giving rise to some of the extreme line ratios observed beyond cosmic noon. Existing models have not yet investigated this possibility in a cosmological emission-line framework. We therefore provide a first implementation that explicitly tests the impact of increased star-cluster masses at high redshift.

As mentioned above, the age of the cluster has a further significant effect on $Q$. However, across galaxy populations with roughly constant star formation in the last 10 Myr, as is typical in ILLUSTRISTNG50, there is no net effect of the age distribution (see also J. Brinchmann et al. 2008). A recent starburst could potentially shift this balance, but we leave an exploration of different star formation histories for future work.

#### 2.2.2 *Increasing ionized gas density* $n_{\rm H}$

An increase in the ionized gas density, $n_{\rm H}$ (roughly equivalent to the electron density $n_{\rm e}$), at fixed volume-filling factor $\epsilon$ leads to denser ionized clumps within the H II region. As shown in equation (1), this effectively reduces the Strömgren radius, producing a more compact H II region and increasing the ionizing photon flux at the ionization front. The result is a higher ionization parameter, as the radiation field becomes more intense relative to the local gas density (X. Liu et al. 2008; F. Bian et al. 2010). Compared to densities of $n_{\rm H} \sim 10$–$10^2$ cm$^{-3}$ locally, nebular gas densities at higher redshift span typical values of $10^2$–$10^3$ cm$^{-3}$, rising to $10^4$–$10^5$ cm$^{-3}$ in compact star-forming regions (M. W. Topping et al. 2025a) and reaching extreme cases of $\gtrsim 10^6$ cm$^{-3}$ in the densest clumps (E. Vanzella et al. 2020). The ionization parameter is likely elevated in these dense environments, which may boost R3 and O32. The exact impact of assuming a certain ionized gas density is line-dependent, as individual transitions are disfavoured when local densities approach their critical values. In this work, we explore LUMEN models with ionized gas densities spanning $n_{\rm H} = 10^2, 10^3$, and $10^4$ cm$^{-3}$ (see Section 6.2.3 for a discussion of the limitations of single-density photoionization models).

#### 2.2.3 *Increasing volume-filling factor* $\epsilon$

The volume-filling factor $\epsilon$ quantifies the fraction of the nebular volume occupied by dense, line-emitting gas. In practice, $\epsilon$ describes how clumpy the ionized medium is, thereby acting as a parametrization of unresolved density variations. A higher $\epsilon$ means that a larger fraction of the nebular volume is filled with dense ionized gas, bringing the volume-averaged gas density closer to $n_{\rm H}$ and increasing $\log U$ according to equation (2), with $\epsilon = 1$ signifying a uniform distribution. Although the physical determinants of $\epsilon$ remain poorly constrained, new high-resolution observations of H II regions, such as those by A. T. Barnes et al. (2026), are beginning to provide empirical estimates. However, since $\epsilon$ cannot exceed unity, its potential to increase $\log U$ is naturally limited. In our modelling, we treat $\epsilon$ as a calibration parameter (see Section 3.3).

### 2.3 Elevated N/O abundance

Elevated N/O abundances have been proposed to explain the offset of $z \sim 2$ galaxies in the N2-BPT diagram (D. Masters et al. 2014; A. E. Shapley et al. 2015; R. L. Sanders et al. 2016). Potential drivers include enrichment by Wolf–Rayet or supermassive stars, the infall of pristine gas, and oxygen-enhanced outflows. While high-$z$ BPT studies have not provided direct evidence for a large systematic increase in N/O across the full galaxy population, recent *JWST* studies suggest that the broader high-$z$ population may exhibit a mild N/O enhancement of order $\sim 0.1$ dex (E. Cataldi et al. 2026). In addition, *JWST* has identified rarer individual systems with super-solar N/O out to $z = 10.6$ (R. Marques-Chaves et al. 2024; I. Morel et al. 2026), demonstrating that much stronger nitrogen enrichment can occur in some high-$z$ galaxies. Using LUMEN, we therefore test to what extent both mild and stronger increases in N/O alone can reproduce the observed N2-BPT offsets.

### 2.4 Leakage of Lyman continuum photons

Leakage of LyC photons from density-bounded regions has been proposed as an alternative explanation for extreme O32 ratios (A. E. Jaskot & M. S. Oey 2013; K. Nakajima & M. Ouchi 2014; A. Plat et al. 2019; Y. I. Izotov et al. 2021; S. R. Flury et al. 2022a, b; A. J. Cameron et al. 2023; S. Mascia et al. 2023; R. L. Sanders et al. 2023b; K. Boyett et al. 2024; A. Calabrò et al. 2024a, b). In a density-bounded H II region, the ionizing radiation is not fully absorbed within the gas cloud, which allows a fraction of ionizing photons to escape into the surrounding medium. This effectively truncates the low-ionization zone. Since the outer layer contributes most of the low-ionization emission, its truncation lowers the [O II]$\lambda\lambda$3727, 3729 line fluxes, while high-ionization lines such as [O III]$\lambda$5007 remain largely unaffected. Unlike an increase in the ionization parameter, this effect primarily reflects the geometric and optical depth structure of the H II region, rather than intrinsic changes in the ionizing source or the gas properties. We account for this changed geometry explicitly by using models for density-bounded H II regions in one of our LUMEN versions.

## 3 THE LUMEN FRAMEWORK

We use LUMEN, our newly developed framework, to test the various physical mechanisms detailed in Section 2. LUMEN is a post-processing methodology providing emission-line predictions from individual H II regions, which are not a priori resolved in cosmological simulations. In this work, we apply our method to the ILLUSTRISTNG50 simulation, but the general LUMEN framework is readily applicable to any hydrodynamic simulation. Future work will explore such extensions. We explain our methodology by focussing on the LUMEN Base model, which was calibrated to reproduce $z = 0$ observations. Descriptions of our entire LUMEN suite, implementing varying ISM and ionizing source conditions for $z > 0$, can be found in Section 3.4. The model suite explores the inclusion of binary stars, massive stars via a higher IMF upper-mass cutoff, $\alpha$-enhanced stellar populations, and AGN contributions (Section 2.1), increased ionization parameters via increased star-cluster masses (Section 2.2.1) and high ionized gas densities (Section 2.2.2), elevated N/O abundances (Section 2.3), LyC leakage from density-bounded H II regions (Section 2.4), and different dust conditions (Section 6.2.5).

### 3.1 The IllustrisTNG50 simulation

Our emission-line predictions are based on ILLUSTRISTNG50 (hereafter TNG50), the highest-resolution volume of the publicly available ILLUSTRISTNG suite of cosmological simulations (A. Pillepich et al. 2018; D. Nelson et al. 2019a). TNG50 was simulated with the moving-mesh code AREPO (V. Springel 2010) in a $(51.7\,\mathrm{cMpc})^3$ volume from $z = 127$ to 0, with dark-matter and gas mass resolutions of $m_{\mathrm{DM}} = 4.5 \times 10^5\,\mathrm{M}_\odot$ and $m_{\mathrm{gas}} = 8 \times 10^4\,\mathrm{M}_\odot$, respectively. The simulation adopts a *Planck* cosmology and includes metal-line cooling, star formation, chemical enrichment, stellar feedback, as well as black-hole seeding, growth, and feedback (Planck Collaboration 2016; R. Weinberger et al. 2017; A. Pillepich et al. 2018). Previous studies have shown that TNG50 reproduces a broad range of galaxy statistics across cosmic time, including stellar mass functions, sizes, colours, star-formation histories, and gas and metal distributions (D. Nelson et al. 2018; A. Pillepich et al. 2018; P. Torrey et al. 2019), making it a suitable basis for our emission-line modelling.

However, several resolution-driven caveats are relevant for this work. TNG50 does not resolve the multiphase ISM, and star-forming gas is instead treated with an effective equation of state above $n_{\mathrm{H}} = 10^{-1}\,\mathrm{cm}^{-3}$ (V. Springel & L. Hernquist 2003). Gas temperatures and densities in this regime should therefore not be interpreted as physical. In addition, stellar particles form stochastically according to a Kennicutt–Schmidt relation and a G. Chabrier (2003) initial mass function, and represent simple stellar populations with masses of order $8 \times 10^4\,\mathrm{M}_\odot$ (R. C. Kennicutt 1998). They therefore cannot capture the full diversity of observed star-cluster masses, nor can realistic H II regions be modelled directly as TNG50 star particles ionizing nearby gas. In Section 3.3, we describe how our post-processing methodology bridges these resolution gaps.

### 3.2 Emission-line models for individual H II regions

In order to obtain nebular emission predictions for individual H II regions, we used CLOUDY (version c13.03; G. J. Ferland et al. 2013) to compute grids of photoionization models ionized by input SEDs from stellar population models. In the Base model, the configuration follows the models from J. Gutkin, S. Charlot & G. Bruzual (2016), which rely on a stellar population synthesis model (G. Bruzual & S. Charlot 2003; Charlot & Bruzual, in preparation) with a standard G. Chabrier (2003) IMF truncated at 0.1 and 100 $\mathrm{M}_\odot$. We fixed the hydrogen density of the ionized gas clumps to $\mathrm{n_H} = 10^2\,\mathrm{cm}^{-3}$ and the dust-to-metal mass ratio to $\xi_d = 0.3$, close to the solar value of 0.36. The CLOUDY calculations were stopped when the electron fraction fell below 0.01 or the gas temperature dropped below 100 K. Otherwise, the models adopted the default convergence criteria of CLOUDY c13.03. The depletion of individual elements onto dust grains follows table 1 in J. Gutkin et al. (2016), which combines the default CLOUDY ISM depletion factors with updates from B. A. Groves, M. A. Dopita & R. S. Sutherland (2004). The adopted solar abundance pattern corresponds to the abundances compiled by A. Bressan et al. (2012), which imply a present-day solar metallicity of $Z_\odot = 0.01524$, a solar interstellar carbon-to-oxygen ratio of $(\mathrm{C/O})_\odot = 0.44$, and a solar nitrogen-to-oxygen ratio of $(\mathrm{N/O})_\odot = 0.07$. Across the entire model grid, we assume a C/O ratio equal to the solar value, as for the optical emission lines discussed in this work, the impact of varying this parameter is negligible (see J. Gutkin et al. 2016).

The resulting grids are parametrized by the age of the stellar population, assuming a single starburst, the ionization parameter at the Strömgren radius $U$, and the interstellar metallicity $Z_{\mathrm{ISM}}$. As stellar populations younger than 10 Myr dominate the output of ionizing photons, we consider Simple Stellar Populations (SSPs) with ages in the range 0–$10^7$ yr, divided into bins of $5 \times 10^5$ yr. The model grids are further parametrized by seven logarithmically spaced values of $\log U$ between $-4$ and $-1$ and by 14 values for the interstellar metallicity: 0.0001, 0.0002, 0.0005, 0.001, 0.002, 0.004, 0.006, 0.008, 0.01, 0.014, 0.017, 0.02, 0.03, and 0.04.

### 3.3 Populating the TNG50 galaxies with H II regions

We used the photoionization models described in Section 3.2 to model H II region emission around individual young star clusters placed in TNG50 galaxies. Fig. 2 illustrates the key steps of this method for one example galaxy at $z = 0$. In short, we identified the young star particles in each TNG50 galaxy (black points, top left), resampled their combined stellar mass into star clusters that reflect the mass diversity of observed star clusters, and reinserted them into the dense gas (black points, bottom left). Assigning stellar population SEDs and H II region photoionization models to each cluster based on metallicity (top middle) and ionization parameter (bottom middle) yields spatially resolved line emission, shown by the H $\alpha$ (top right) and [O III]$\lambda$5007 (bottom right) surface-brightness maps. The methodology is described in more detail below.

#### 3.3.1 *Resampling young stellar mass into individual star clusters*

As part of the resampling procedure, we sorted all star particles younger than 10 Myr into age bins between 0 and $10^7$ yr, spaced by $5 \times 10^5$ yr. In each age bin, we computed the total stellar mass formed by summing up the masses of all contained star particles. We then took advantage of a result from the GRIFFIN dwarf merger simulation (N. Lahén et al. 2020), which models the cloud collapse and the subsequent star cluster formation self-consistently at sub-parsec resolution. N. Lahén et al. (2020) determined a cluster mass function (CMF), in which the number of clusters $dN$ per unit mass $dM$ follows a power law, $dN/dM \propto M^{\alpha}$. For clusters above $10^3$ M$_\odot$, their best-fit index is $\alpha = -1.96 \pm 0.18$. This value is in excellent agreement with observational determinations of $\alpha \approx -2$ across clusters embedded in nearby molecular clouds (C. J. Lada & E. A. Lada 2003), dwarf galaxies (D. O. Cook et al. 2012), discs of nearby galaxies (A. Bik et al. 2002; N. McCrady & J. R. Graham 2007; S. S. Larsen 2009; A. Mok, R. Chandar & S. M. Fall 2019; M. R. Krumholz et al. 2019), galaxy mergers (Q. Zhang & S. M. Fall 1999), as well as for high-$z$ galaxies (M. Dessauges-Zavadsky & A. Adamo 2018; A. Claeyssens et al. 2025, 2026). N. Lahén et al. (2020) further found that the maximum cluster mass $M_{\rm cl,max}$ is set by the star formation surface density $\Sigma_{\rm SFR}$. We thus determined the $M_{\rm cl,max}$ using an estimate of the galaxies' $\Sigma_{\rm SFR}$ within the half-mass radius. For each galaxy, we resampled the young stellar mass in each age bin using the CMF with $\alpha = -1.96$ between $10^3$ M$_\odot$ and $M_{\rm cl,max}$. Due to the relatively steep power-law slope, the exact value of $M_{\rm cl,max}$ adopted for a galaxy does not affect the results significantly, since clusters above $10^5$ M$_\odot$ are relatively rare when assuming a minimum mass of $10^3$ M$_\odot$. The impact of systematically more massive clusters existing at higher redshift will be considered in Section 5.

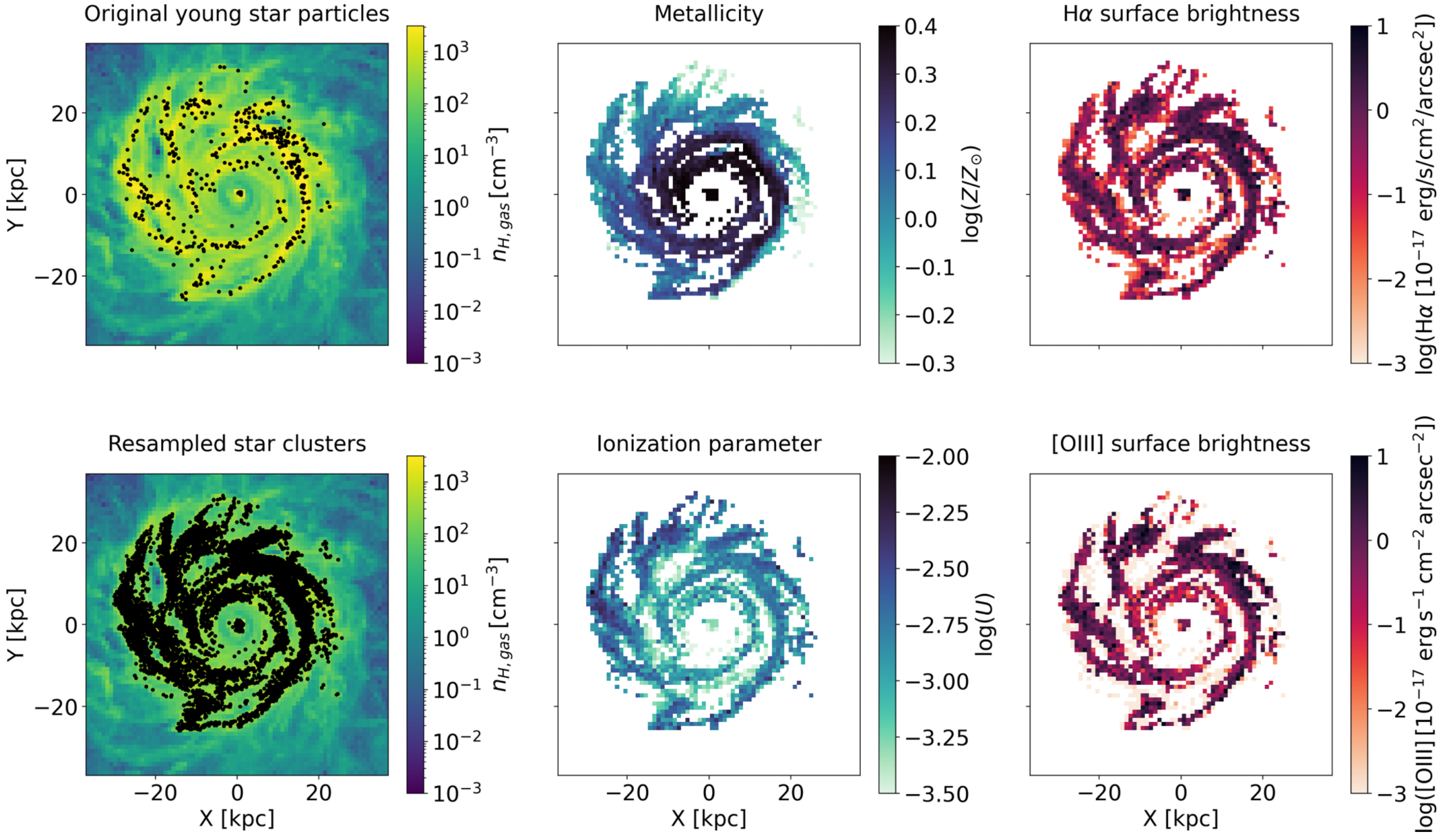

**Figure 2.** LUMEN method illustrated for one example galaxy at $z = 0$ in the Base model. The young star particles formed in IllustrisTNG (black points, top left) are resampled into young star clusters (black points, bottom left) and reinserted into the dense gas. Based on the metallicity (top middle) and the ionization parameter (bottom middle), photoionization models coupled to each cluster predict line emission from spatially distributed H II regions, shown here as H $\alpha$ (top right) and [O III]$\lambda$5007 (bottom right) surface-brightness maps.

Overall, we obtained a population of star clusters that follows the CMF and retains the roughly constant age distribution set by the original star formation history of each galaxy. These star clusters provide the ionizing radiation for each H II region, as explained in Sections 3.3.2 and 3.3.3.

### 3.3.2 *Placing star clusters into the dense gas*

The star clusters were then reassigned to dense gas cells within each simulated galaxy. Eligible cells were defined as star-forming gas with $n_{\rm H,gas} > 10^{-1}$ cm$^{-3}$, following TNG50, that lie within the galactocentric radius of the most distant young star particle. We then randomly selected one gas cell per cluster from this sample.

The coupling procedure of the star clusters to the gas was informed by results from high-resolution simulations and observations. Detailed modelling has shown that new clusters form in the densest gas, but early stellar feedback from photoionization, radiation pressure, and far-ultraviolet heating disrupts their surroundings. Within a few Myr, expanding H II regions and photoevaporative flows can clear the immediate environment and reduce the local gas density (T. Peters et al. 2017; J.-G. Kim, W.-T. Kim & E. C. Ostriker 2018; S. Haid et al. 2019). As a result, the youngest clusters are expected to remain at least partially embedded in dense gas, whereas clusters aged $\gtrsim$ 5–10 Myr are typically displaced from their natal material and reside in more diffuse environments created by feedback-driven clearing. Observations paint a similar picture. Young stellar objects and newly formed clusters ($<$ 5 Myr) are strongly concentrated in the densest filaments and ridges of molecular clouds (C. J. Lada & E. A. Lada 2003; S. T. Megeath, R. A. Gutermuth & M. A. Kounkel 2022), and the cloud lifecycle framework emphasizes that the highest-density phases set the initial conditions for cluster formation (M. Chevance et al. 2020). Older clusters (5–10 Myr) show weaker spatial correlations with dense gas as feedback disperses their birth sites.

In order to capture the tight relationship between age and environmental density in the earliest evolutionary stages, we ranked star clusters by age and gas cells by density, assigning the youngest clusters to the densest gas cells (corresponding to typical H II-region densities $\gtrsim 10^2\,\mathrm{cm^{-3}}$), while older clusters were placed in progressively more diffuse environments ($\lesssim$ 1–10 $\mathrm{cm^{-3}}$). This scheme does not explicitly follow the time-evolving displacement of clusters from their birth clouds. However, because the ionizing photon rate $Q$ declines steeply within the first few Myr (G. Bruzual & S. Charlot 2003), the nebular emission is dominated by the youngest clusters. Therefore the precise placement of older clusters does not strongly affect our results.

For each star cluster, the gas density, $n_{\mathrm{H,gas}}$, and metallicity, $Z_{\mathrm{gas}}$, of its assigned gas cell were adopted as the properties of the surrounding H II region. When the number of clusters exceeded the number of eligible gas cells, we allowed multiple clusters to be assigned to the same gas cell. We considered each of these to be its own independent H II region, treating them as two neighbouring, but not overlapping, regions with the same gas properties illuminated by different star clusters. This does not imply that the same gas is ionized several times. Alternative treatments, including limiting the number of star clusters to the number of eligible gas cells or combining star clusters assigned to the same cell into a single H II region, had no significant effect on the emission-line predictions.

#### *3.3.3 Coupling emission-line models to each star cluster*

Photoionization models were assigned based on the gas-phase metallicity $Z_{\mathrm{gas}}$ and an estimate of the ionization parameter $U$. We computed $U$ using equation (2) with an assumed ionized gas density $n_{\mathrm{H}} = 10^2\,\mathrm{cm^{-3}}$, the ionizing photon rate $Q$, and the volume filling factor $\epsilon$. To derive $Q$ for each cluster, we matched a G. Bruzual & S. Charlot (2003) SSP model according to the age of the star cluster and the gas-phase metallicity, assuming $Z_\star = Z_{\mathrm{gas}}$. The SED was scaled by the cluster mass and integrated up to the Lyman limit ($\lambda_{\mathrm{ly}} = 911.3$ Å) to obtain $Q$. The volume filling factor is commonly estimated as the ratio of the volume-averaged gas density in the ionized region to $n_{\mathrm{H}}$. However, as noted in Section 3.1, individual gas cell densities are not reliable tracers of physical ISM conditions.

We therefore treated $\epsilon$ as a calibration parameter to reproduce an empirically motivated distribution of ionization parameters. Observations consistently show that lower-metallicity gas exhibits higher ionization parameters (e.g. T. Nagao, R. Maiolino & A. Marconi 2006; M. A. Dopita et al. 2006a, b; E. Pérez-Montero 2014; D. Carton et al. 2017; R. L. Sanders et al. 2020). Inspired by D. Carton et al. (2017), we adopted $\log U = -\log(Z_{\mathrm{gas}}/Z_\odot) - 3.25$ to estimate $\log U$ from the mean gas-phase metallicity $Z_{\mathrm{gas}}$ in 50 annuli spanning the galaxy centre to the outermost young star particle. Using the mean $Q$ within each annulus, we then inverted equation (2) to derive a radial profile of $\epsilon$. Values exceeding unity were capped at $\epsilon = 1$. Varying the number of annuli did not significantly affect the inferred $\epsilon$ profiles or any subsequent results.

Based on the calibrated $\epsilon$ values, we computed the individual $U$ (equation 2) and Strömgren radius $R_{\mathrm{S}}$ (equation 1) for each star cluster using its own $Q$. Due to this calibration, we recover ionization parameter gradients that are on average anticorrelated with the metallicity gradient, but contain variation introduced by the properties of the star clusters (see middle panels in Fig. 2). Since the youngest star clusters, which produce the highest $Q$, were placed in the densest gas cells, which in TNG50 are usually the most metal-rich, the resulting cluster-to-cluster variation can weaken or even invert the initially positive ionization-parameter gradient.

We then used $\log U$, alongside the star cluster age and $Z_{\mathrm{gas}}$, to match each cluster to the closest entry in the emission-line model grid. Applying this methodology to galaxies in TNG50 snapshots spanning redshifts $z = 0$–8, we obtained emission-line predictions for a diverse population of individual H II regions across cosmic time. As the line ratios considered here are constructed from closely spaced lines and are therefore only weakly affected by diffuse ISM attenuation, we do not apply such attenuation to the predicted emission-line strengths. In addition to the full TNG50 snapshots, we included 50 galaxies from ILLUSTRISTNG100, resimulated as zooms at TNG50 resolution as part of the BLACKDAWN simulation suite (Farcy et al., in preparation). The targeted galaxies were among the most massive ones at $z = 3$, thus we effectively extended our mass sampling beyond the limited high-mass coverage of the TNG50 volume.

### 3.4 Model variations for $z > 0$

Table 2 shows an overview of the LUMEN suite of models we employ to represent the varying ionizing source and gas conditions for $z > 0$ described in Section 2. All LUMEN variants are defined relative to the Base model, which was calibrated to reproduce local H II region and galaxy emission-line properties. Each model modifies a specific subset of parameters while holding the remaining framework fixed. We note that the variations explored here represent intentionally extreme cases, as the parameter changes are applied uniformly across all galaxies. These models are therefore not meant to reflect fully realistic galaxy populations, but rather to illustrate the global direction and magnitude of emission-line shifts that each physical mechanism may produce.

#### *3.4.1 Variations modelling harder ionizing radiation*

A range of physical processes can harden the ionizing radiation, either related to the stellar population or additional non-stellar sources, such as AGNs. We describe the corresponding model variations below.

(i) Binary stars (BPASS binary): Instead of SSP SEDs from G. Bruzual & S. Charlot (2003), we used models from Binary Population And Spectral Synthesis v2.2.1 (BPASS; J. J. Eldridge et al. 2017) with ISM metallicities of 0.0001, 0.001, 0.002, 0.004, 0.006, 0.008, 0.010, 0.014, 0.020, 0.030, and 0.040. BPASS accounts for the effects of binary star interactions, such as mass transfer and mergers, which result in harder ionizing spectra via hotter stars and prolong the production of ionizing photons beyond 10 Myr (e.g. L. Xiao, E. R. Stanway & J. J. Eldridge 2018; M. Lecroq et al. 2024). The impact of binary stars is strongest at older stellar ages and in high-ionization lines such as He II. In the N2-BPT, S2-VO87, and O32–R23 diagrams, the differences between single- and binary-star models are modest (see Figs 7, 8, and 9), as the relevant lines are less sensitive to photons above the He II-ionizing threshold. Furthermore, the older-age binary phase contributes relatively little for realistic star-formation histories. The BPASS binary SEDs additionally differ from the updated G. Bruzual & S. Charlot (2003) single-star SEDs through the underlying stellar tracks, atmospheres, and treatment of Wolf–Rayet stars. The lower Wolf–Rayet contribution in BPASS at low metallicity offsets most of the remaining hardening from binary evolution in the diagnostic diagrams considered here. As a result, the BPASS-binary predictions appear similar to the Base model.

**Table 2.** Column (1) Stellar population model used for the SED (CB = single star models from Charlot & Bruzual, in preparation; BPASS = Binary Population And Spectral Synthesis, J. J. Eldridge et al. 2017). Column (2) IMF upper-mass cutoff in $M_{\odot}$. Column (3) Stellar metallicity used in the SED. Column (4) Radiative sources included (SF = stellar; AGN = AGN narrow-line region). Column (5) Minimum cluster mass in $M_{\odot}$. Column (6) Ionized hydrogen density in $cm^{-3}$. Column (7) Nitrogen-to-oxygen abundance relative to the J. Gutkin et al. (2016, G16) N/O–O/H relation. Column (8) Lyman-continuum escape fraction. Column (9) Dust-to-metal mass ratio. Column (10) Dust grain size distribution in CLOUDY.

| Model | SED | $m_{up}$ | $Z_*$ | Source | $\log M_{cl,min}$ | $\log n_H$ | N/O | $f_{esc}$ | $\xi_d$ | Grains |
|---|---|---|---|---|---|---|---|---|---|---|
| Base | CB | 100 | $Z_{gas}$ | SF | 3 | 2 | G16 | 0 | 0.3 | ISM |
| BPASS | BPASS | 100 | $Z_{gas}$ | SF | 3 | 2 | G16 | 0 | 0.3 | ISM |
| High-$m_{up}$ | CB | 300 | $Z_{gas}$ | SF | 3 | 2 | G16 | 0 | 0.3 | ISM |
| $\alpha$-enhanced | CB | 100 | $Z_{gas}/5$ | SF | 3 | 2 | G16 | 0 | 0.3 | ISM |
| High-$m_{up}$ + $\alpha$-enhanced | CB | 300 | $Z_{gas}/5$ | SF | 3 | 2 | G16 | 0 | 0.3 | ISM |
| Weak/strong AGNs | CB | 100 | $Z_{gas}$ | SF+AGN | 3 | 2 | G16 | 0 | 0.3 | ISM |
| High star-cluster masses | CB | 100 | $Z_{gas}$ | SF | 5, 6 | 2 | G16 | 0 | 0.3 | ISM |
| High ionized gas density | CB | 100 | $Z_{gas}$ | SF | 3 | 3, 4 | G16 | 0 | 0.3 | ISM |
| Elevated N/O | CB | 100 | $Z_{gas}$ | SF | 3 | 2 | G16 + 0.1, 0.6 | 0 | 0.3 | ISM |
| LyC leakage | CB | 100 | $Z_{gas}$ | SF | 3 | 2 | G16 | 0.2, 0.5 | 0.3 | ISM |
| Low dust-to-metal | CB | 100 | $Z_{gas}$ | SF | 5, 6 | 2 | G16 | 0 | 0.1 | ISM |
| Orion grains | CB | 100 | $Z_{gas}$ | SF | 5, 6 | 2 | G16 | 0 | 0.3 | Orion |
| 'Master' | CB | 300 | $Z_{gas}/5$ | SF | 5 | 2 | G16 + 0.1 | 0 | 0.3 | ISM |

(ii) High IMF upper-mass cutoff ($m_{up} = 300 M_{\odot}$): We increased the upper limit of the stellar initial mass function from 100 to 300 $M_{\odot}$, resulting in harder ionizing radiation. Variations in the IMF slope (e.g. a top-heavy IMF) could further enhance the production of high-energy photons by increasing the relative number of massive stars. While such IMF modifications have been explored in the context of high-$z$ galaxies and extreme star-forming environments (C. M. Baugh et al. 2005; P. G. Dokkum 2008; M. L. Gunawardhana et al. 2011; M. Marks et al. 2012; D. Narayanan & R. Davé 2012), we do not vary the IMF slope in this work, as current observational and theoretical constraints on such variations remain limited and their effects are degenerate with other model parameters, such as star formation history, metallicity, and stellar evolution prescriptions.

(iii) Stellar $\alpha$-enhancement: Stellar $\alpha$-enhancement refers to supersolar $\alpha$/Fe ratios in the stellar population at fixed metallicity. These conditions are not natively found in TNG50. Following the reasoning of C. C. Steidel et al. (2016), we therefore approximate their effect in post-processing by decoupling the stellar and gas-phase metallicities. To first order, the stellar opacity and mass-loss rates that control the hardness of the ionizing spectrum depend mostly on stellar Fe/H (see also $\alpha$-enhanced models by C. M. Byrne et al. 2022), while the nebular emission lines are mainly affected by the gas-phase O/H abundance. Using a lower stellar metallicity at fixed gas-phase metallicity therefore approximates the harder ionizing spectra expected from $\alpha$-enhanced stellar populations, while keeping the nebular oxygen abundance fixed. We implement this by reducing the stellar metallicity of the SEDs to $Z_{\star} = Z_{gas}/5$, reflecting the supernova nucleosynthesis yields from K. Nomoto et al. (2006). This corresponds to a relatively extreme level of stellar $\alpha$-enhancement compared to recent observational constraints at cosmic noon, which typically find offsets closer to $Z_{gas} \simeq 2$–$3\,Z_{\star}$ (A. L. Strom et al. 2022; T. M. Stanton et al. 2024). We therefore interpret this model as an intentionally strong limiting case for the impact of stellar $\alpha$-enhancement on the ionizing spectrum, rather than as a representative abundance pattern for all high-$z$ galaxies.

(iv) High IMF upper-mass cutoff and stellar $\alpha$-enhancement: We constructed an additional model combining the spectral hardness from an IMF upper-mass cutoff of 300 $M_{\odot}$ and stellar $\alpha$-enhancement.

(v) Weak and strong AGNs: In addition to the stellar component, we included AGN narrow-line region (NLR) contributions following M. Hirschmann et al. (2023b). For this, we used NLR photoionization models from A. Feltre, S. Charlot & J. Gutkin (2016), which were updated to include the inner radius of the NLR gas distribution and the internal microturbulence velocity of the emitting clouds as additional free parameters (M. Mignoli et al. 2019). Every galaxy was assigned an NLR model by matching the mass-weighted gas-phase metallicity and ionization parameter $U$ within a 1 kpc sphere centred on the black hole. We distinguished between weak and strong AGNs using the relative strength of the black-hole accretion rate (BHAR) and the SFR, with a threshold value of BHAR/(BHAR + SFR) = $10^{-2.5}$.

### *3.4.2 Variations modelling an increase in the ionization parameter*

Due to the assumed anticorrelation of the ionization parameter and the metallicity and generally decreasing metallicities with increasing redshift, the Base model already contains an implicit increase in ionization parameter. However, as we will show in Section 5, this effect is not strong enough to explain most of the high-$z$ observations in the N2-BPT, S2-VO87, and R23–O32 diagrams. We include the following model versions that further increase $\log U$:

(i) Increased star-cluster masses ($\log M_{cl,min} = 5, 6$): As described in Section 2, we hypothesize that the high star-cluster masses observed in high-$z$ galaxies increase the rate of ionizing photons $Q$ and thereby the ionization parameter. The shape of the cluster mass function at $z > 0$ remains uncertain, with some evidence that compact high-$z$ systems may require either a flatter mass-function slope or a higher low-mass cutoff (E. Vanzella et al.

2026). Current observations of the high-mass end remain broadly consistent with a canonical $\alpha \simeq -2$ power law, as found for resolved high-$z$ star clusters by A. Claeyssens et al. (2026) and unresolved stellar clumps (M. Dessauges-Zavadsky & A. Adamo 2018; A. Claeyssens et al. 2025). In order to test the impact of systematically higher star-cluster masses at high redshift, we therefore do not modify the slope of the cluster mass function, but instead test the effect of shifting the minimum star-cluster mass towards higher masses. We therefore resampled the young stellar mass in TNG50 using the same mass function as in Section 3.3.1, but with a minimum star-cluster mass $M_{\rm cl,min}$ of $10^5$ M$_\odot$ and $10^6$ M$_\odot$, respectively. For the version with $10^6$ M$_\odot$, we widened the age-bin width to $2.5 \times 10^6$ yr to retain sufficient young stellar mass to draw at least one cluster per age bin in most galaxies, resulting in bin edges of $[0, 2.5, 5, 7.5, 10] \times 10^6$ yr.

(ii) High ionized gas density ($\log n_{\rm H} = 3, 4$): We increased the ionized hydrogen gas density from $10^2$ to $10^3$ and $10^4$ cm$^{-3}$. This not only increases the ionization parameters, but also has more complex effects on individual line strengths, as discussed in Section 2.

Lastly, we adopt the roughly flat stellar age distributions from the simulation and do not vary the recent star formation histories in our models. Exploring the impact of generally younger ages due to bursty star formation will be the focus of future work.

#### *3.4.3 Variations modelling elevated N/O abundances*

The TNG50 simulation does not intrinsically exhibit strong variations in N/O, as the subgrid chemical enrichment model does not track all processes possibly responsible for nitrogen enhancement (e.g. secondary production or localized enrichment from Wolf–Rayet stars). To still explore the effect of enhanced N/O, we instead model the resulting effects by shifting the N/O–O/H abundance relation (J. Gutkin et al. 2016, equation 11) relative to our Base model.

(i) Mildly and strongly elevated N/O: We increased the N/O abundance by 0.1 and 0.6 dex at fixed O/H, representing mild and strong N/O enhancement, respectively. We note that while we call the +0.6 dex version strongly N/O-enhanced within the context of our models, this variation agrees with the upper end of the scatter in the local N/O–O/H relation from D. C. Nicholls et al. (2017) and is well below the highest observed abundances of known nitrogen emitters at high redshift (R. Marques-Chaves et al. 2024; I. Morel et al. 2026).

#### *3.4.4 Variation modelling Lyman continuum leakage*

In addition to radiation-bounded models, in which all ionizing photons are absorbed within the H II region, we included a model variant representing density-bounded conditions.

(i) Lyman continuum leakage from density-bounded H II regions ($f_{\rm esc} = 0.2, 0.5$): We made use of the density-bounded H II regions modelled with CLOUDY by A. Plat et al. (2019). In these models, the photoionized gas is truncated at a fixed hydrogen column density corresponding to a prescribed Lyman-continuum escape fraction $f_{\rm esc}$. We used an escape fraction of $f_{\rm esc} = 0.2$, such that roughly 20 per cent of the ionizing photons escape the H II region, consistent with expectations for galaxies contributing to reionization (R. P. Naidu et al. 2020). Additionally, we included another model with $f_{\rm esc} = 0.5$ to assess the emission-line signatures of more extreme LyC leakers.

#### *3.4.5 Variations modelling different dust characteristics*

Different dust treatments can further affect the absorption and re-emission of radiation, as well as the depletion of metals onto dust grains. In Section 6, we explore the impact of two alternative CLOUDY dust prescriptions on our emission-line predictions. To illustrate the impact of these changes, we applied them to models with high star-cluster masses, where the effect is particularly strong.

(i) Lower dust-to-metal ratio ($\xi_d = 0.1$): We decreased the dust-to-metal mass ratio $\xi_d$ from 0.3 to 0.1.

(ii) Orion-like dust grains: We adopted an Orion-like grain size distribution in place of the Milky Way ISM-based distribution used in other LUMEN models (see section 5.4 in G. J. Ferland et al. 2017). The Orion-like distribution contains fewer small dust grains, resulting in less dust absorption of ionizing photons, which strengthens the Balmer lines.

#### *3.4.6 Composite model: the 'Master' version for high-z*

Lastly, in order to reproduce all high-$z$ constraints simultaneously, we constructed one composite LUMEN version, hereafter the 'Master' model.

(i) Master model: We combined spectral hardness from (i) an IMF upper-mass cutoff of 300 M$_\odot$ and (ii) stellar $\alpha$-enhancement, with (iii) star-cluster masses $> 10^5$ M$_\odot$, and (iv) mildly elevated N/O abundances.

## 4 VALIDATION OF BASE MODEL AT $Z = 0$

Before proceeding with our main analysis, we first validate the LUMEN methodology by comparing emission-line predictions from our Base model against observational data for nearby galaxies. Section 4.1 shows that the modelled H II regions reproduce observed properties, and Section 4.2 demonstrates that integrated galaxy emission matches local optical line-ratio diagrams and reproduces strong-line calibrations for SFRs and metallicity. This successful validation enables us to extend LUMEN to the high-$z$ regime in Section 5.

### 4.1 Resolved properties

#### *4.1.1 H II region properties*

Early studies resolving individual H II regions in nearby galaxies (R. C. Kennicutt, F. Bresolin & D. R. Garnett 2003; E. Rosolowsky & J. D. Simon 2007), and later the CHemical Abundances Of Spirals project (CHAOS, D. A. Berg et al. 2015; K. V. Croxall et al. 2015, 2016), have revealed substantial diversity in nebular conditions. More recent integral-field spectroscopy with MUSE has extended this work to larger samples and fainter regions (K. Grasha et al. 2022; F. Santoro et al. 2022; A. Feltre et al. 2026; R. Garner et al. 2026). The most comprehensive compilation to date comes from the PHANGS-MUSE survey, which identified over 23 000 H II regions across 19 nearby galaxies (F. Santoro et al. 2022). Subsequent analyses (E. Congiu et al. 2023; B. Groves et al. 2023; D. Pathak et al. 2025; A. T. Barnes et al. 2026) expanded on this work by refining nebular classifications, quantifying environmental trends, and linking nebular conditions to stellar populations and feedback. Together, these studies have established a detailed reference sample of resolved H II regions of nearby galaxies, which we used to calibrate our model.

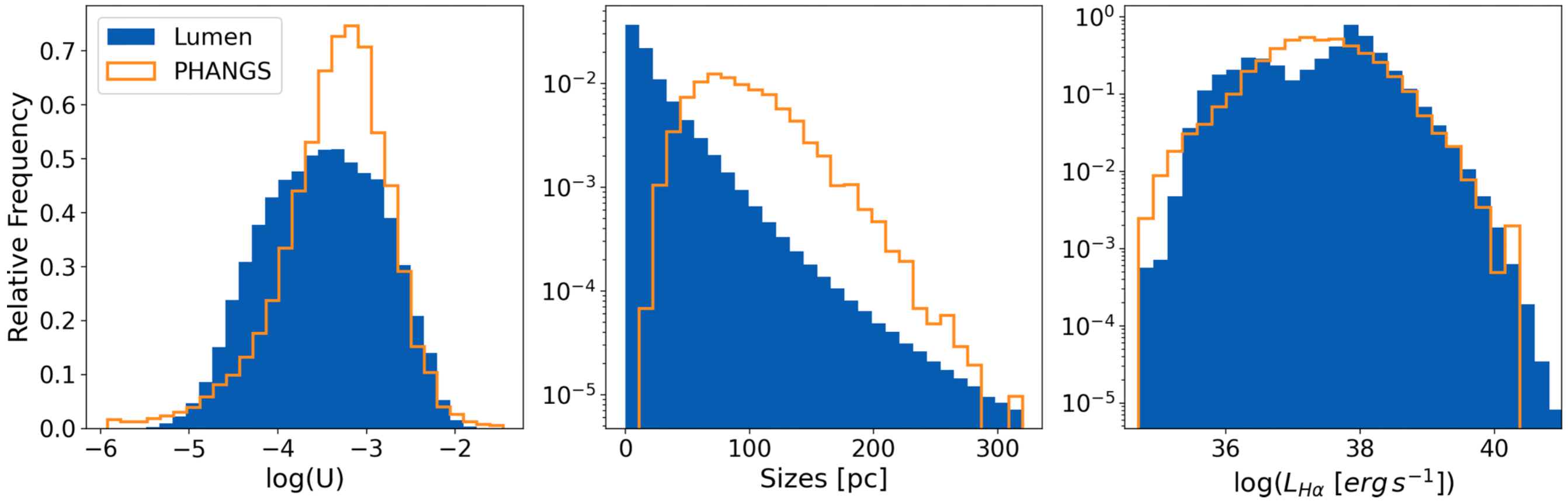

**Figure 3.** Comparison of physical H II region properties from the PHANGS catalogue (orange) and the LUMEN $z = 0$ sample (blue). Shown are the ionization parameters (left), circularized PHANGS radii and LUMEN Strömgren radii (middle), and H $\alpha$ luminosities (right).

Fig. 3 shows a comparison of the physical H II region properties from the PHANGS catalogue (orange) and the LUMEN $z = 0$ sample (blue). From left to right, we show the normalized distribution of the ionization parameters, the circularized radii from PHANGS and the estimated Strömgren radii from LUMEN, as well as the H $\alpha$ luminosities. We note that a strict one-to-one comparison between H II regions in matched PHANGS and LUMEN galaxies was not possible due to the low gas-phase metallicities that were determined for PHANGS, which generally lie below the local mass–metallicity relation (around $12 + \log(\mathrm{O/H}) = 8.3$–$8.6$ at $M_\star = 10^9$–$10^{11}$ $\mathrm{M}_\odot$; see B. Groves et al. 2023). At fixed metallicity, TNG50 galaxies have significantly higher stellar masses. We therefore compare the full PHANGS and LUMEN samples to determine whether the model captures the observed H II region properties.

The ionization-parameter and H $\alpha$-luminosity distributions are remarkably consistent between the two samples. The ionization parameters span $\log U = -5.9$ to $-1.5$, with a slightly broader distribution in the LUMEN sample. H $\alpha$ luminosities range from $\log L_{\mathrm{H}\alpha} = 34.7$ to $\sim 40.5\,\mathrm{erg\,s^{-1}}$. The inferred H II region sizes show comparatively larger differences, which reflect their stronger dependence on the assumed gas density and geometry. In LUMEN, Strömgren-sphere estimates yield a median radius of $\approx 14$ pc, while E. Congiu et al. (2023) report a median circularized radius of $\approx 88$ pc based on segmentation of MUSE maps. This difference likely arises from limitations in spatial resolution and sampling. Observationally, small nebulae will either blend into larger complexes or be isolated and too faint to be observed. This is supported by A. T. Barnes et al. (2026), who cross-matched 4882 PHANGS-MUSE regions with high-resolution HST imaging and identified substructures down to the 10 pc resolution limit (median $\approx 20$ pc), which yielded systematically smaller H II region sizes.

LUMEN also predicts a few H II regions larger than 300 pc (0.06 per cent predicted across 6627 galaxies, not shown in Fig. 3). These are likely underrepresented in PHANGS because they are intrinsically rare, artificially fragmented into multiple subregions, or unobserved due to potentially low densities and faint surface brightnesses. Nevertheless, their sizes are consistent with the upper end of observed H II region distributions. In nearby galaxies, H II regions can reach sizes of several hundred pc (R. C. Kennicutt & N. J. Evans 2012), with the largest examples extending to $\sim 600$ pc (K. Grasha et al. 2022) and, in some cases, approaching $\sim 1$ kpc (L. K. Hunt & H. Hirashita 2009). Overall, we conclude that LUMEN produces physically plausible H II region properties at $z = 0$, well-aligned with those observed in nearby galaxies.

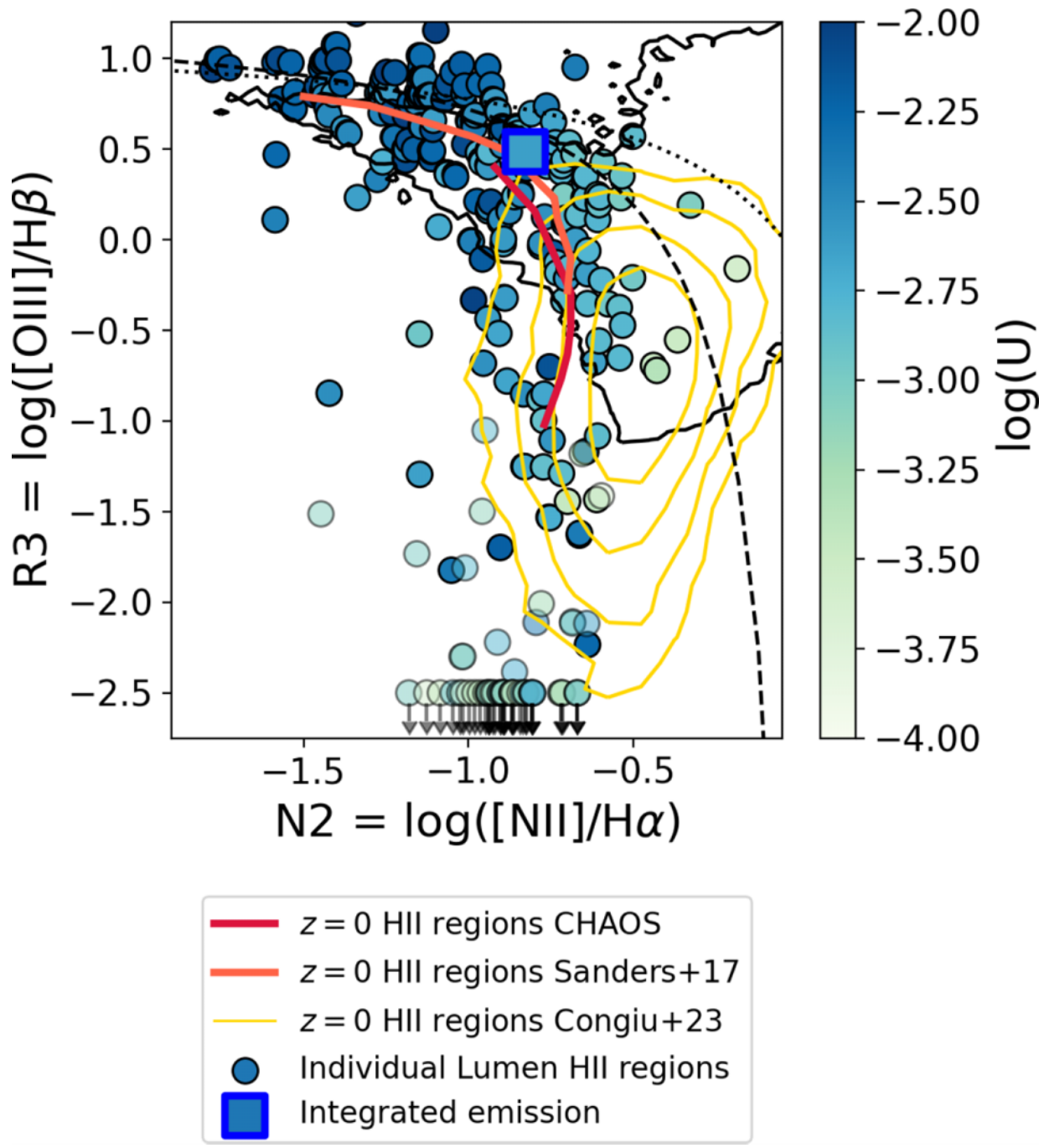

**Figure 4.** Location of LUMEN H II regions in one example galaxy on the N2-BPT diagram. Shown are all H II regions (circular markers) and their integrated emission across the entire galaxy (square marker) coloured by the ionization parameter. H II regions with emission lines below $10^{-20}$ erg s$^{-1}$ cm$^{-2}$ are shown with reduced opacity. Further included are local SDSS galaxies (black outline), empirical selection criteria (black lines) and the mean of observed $z = 0$ H II regions from R. L. Sanders et al. (2017; orange) and CHAOS (D. A. Berg et al. 2015; K. V. Croxall et al. 2015; red), as well as contours from PHANGS (E. Congiu et al. 2023; yellow).

### 4.1.2 *H II regions in the BPT*

Fig. 4 shows the location of individual H II regions from the LUMEN galaxy in Fig. 2 on the N2-BPT diagram. Circular markers represent individual H II regions, colour-coded by their ionization parameters, while the square marker indicates the integrated emission over the entire galaxy. H II regions where at least one emission line lies below a flux cut of $10^{-20}$ erg s$^{-1}$ cm$^{-2}$ are shown with reduced opacity. We further show local SDSS galaxies (black outline), the mean loci of observed $z = 0$ H II regions from CHAOS (D. A. Berg et al. 2015; red), and R. L. Sanders et al. (2017; orange), as well as contours from the PHANGS sample (E. Congiu et al. 2023; yellow). All three of these samples occupy distinct locations in the N2-BPT diagram, with R. L. Sanders et al. (2017) covering the high R3 tail of the star-forming branch, the CHAOS regions spanning intermediate R3 values at relatively low N2, and the PHANGS sample clustering around low-intermediate R3 and slightly higher N2 values. These offsets likely reflect a combination of aperture and spatial resolution effects, differences in stellar continuum subtraction and extinction correction, as well as intrinsic variations in metallicity and ionization parameter across the samples.

The modelled LUMEN H II regions span the full parameter space covered by these observations and exhibit a broad range of ionization parameters, consistent with diverse excitation conditions within a single galaxy. The subset of H II regions shown with downward-pointing arrows lies at significantly lower R3, largely below the $10^{-20}$ erg s$^{-1}$ cm$^{-2}$ flux limit. Overall, Fig. 4 illustrates that global line ratios reflect the combined output of many H II regions, and that local variations likely impact inferred physical properties (X. Ji et al. 2024; C. Morisset et al. 2025; M. Usui et al. 2025; D. A. Berg et al. 2026; A. P. Vijayan et al. 2026).

## 4.2 Integrated properties

For the following analysis, we computed the total integrated line luminosities per galaxy by summing up the emission-line contributions from all H II regions. Other H II region properties, such as the gas-phase metallicities and ionization parameters, were summarized as H $\alpha$ luminosity-weighted averages to represent the properties of the entire galaxy.

### 4.2.1 *Line-ratio diagrams at $z = 0$*

Fig. 5 shows the integrated emission from the $z = 0$ LUMEN Base model in the N2-BPT diagram (top left), the VO87 diagrams, R3 versus S2 (top right), R3 versus [O I]$\lambda$6300/H $\alpha$ (bottom left), and the O32–R23 diagram (bottom right). For reference, we include the theoretical upper starburst limits from L. J. Kewley et al. (2001a; dotted lines) in the N2-BPT and VO87 diagrams, and the empirical AGN lower boundary from G. Kauffmann et al. (2003; dashed line) in the N2-BPT diagram.

We distinguish between galaxies with $M_\star > 10^8\,\mathrm{M}_\odot$ (dark blue contours), used in the main analysis, and an extended sample down to $10^7\,\mathrm{M}_\odot$ (light blue shading). The N2-BPT diagram also shows the corresponding medians (dark and light blue lines). To match the SDSS sample (black outline, red dashed median), we apply an approximate luminosity limit of $10^{39}$ erg s$^{-1}$ (K. I. Caputi et al. 2011) to the $M_\star > 10^8\,\mathrm{M}_\odot$ population. The $M_\star > 10^7\,\mathrm{M}_\odot$ sample is shown without this cut to illustrate the full distribution. Galaxies with stellar masses between $10^7$ and $10^8\,\mathrm{M}_\odot$ are intrinsically faint and lie generally close to or below the SDSS low-luminosity limit. While these galaxies may be considered resolved for some purposes (see D. Nelson et al. 2019a), they are represented by merely 125–1250 star particles. At these resolutions, the emission-line predictions are less reliable and we exclude them from the analysis in Section 5.

LUMEN galaxies from the Base model occupy the respective star-forming branches in the N2-BPT and VO87 diagrams, thus showing good qualitative agreement with observations. In the N2-BPT diagram, the main population above the $10^8\,\mathrm{M}_\odot$ cut exhibits a median sequence that lies closer to the G. Kauffmann et al. (2003) criterion. This is because at masses below $10^{9.5}\,\mathrm{M}_\odot$, the TNG50 metallicities are on average higher than the SDSS metallicities (see also P. Torrey et al. 2019). Extending the sample down to $10^7\,\mathrm{M}_\odot$ and removing the luminosity limit introduces systems with lower metallicities, which in turn shifts the median towards slightly lower R3 values.

We further note that the contribution from unmodelled diffuse ionized gas (DIG), characterized by low ionization states, would likely move the LUMEN galaxies towards higher S2 and [O I]$\lambda$6300/H $\alpha$ ratios (R. L. Sanders et al. 2017; F. Mannucci et al. 2021; E. Congiu et al. 2023). However, we consider this effect to be negligible for galaxies with high specific SFR (sSFR) close to the starburst criterion, as the H II regions occupy a large volume of the ionized gas and outshine any DIG background emission (R. L. Sanders et al. 2017; F. Mannucci et al. 2021). As we focus our main analysis in Section 5 on the gap between the most star-forming local galaxies and the observed high-$z$ galaxies, which due to high sSFR and $\Sigma_{\mathrm{SFR}}$ should also be relatively compact, we consider the DIG contribution negligible for our purposes.

Lastly, we examine the O32–R23 diagram. Unlike in the other diagnostic diagrams, where we show the full SDSS sample and use the local classification lines as a visual reference, we apply the N2-BPT criterion from G. Kauffmann et al. (2003) to remove likely AGNs from the catalogue. LUMEN reproduces the resulting sample well, except for the high-R23, low-O32 tail. This is because R23 initially increases with metallicity at low metallicities, but declines at higher metallicities as efficient metal-line cooling lowers the electron temperature and suppresses collisional excitation, which weakens the oxygen lines. In the Base model, the photoionization predictions from our H II region models saturate at R23 $\sim 0.8$ for O32 $< -0.5$, and therefore do not extend to the most extreme R23 values in the SDSS sample. One possible explanation for this slight mismatch is residual AGN contamination, which may boost observed R23 ratios. Past modelling work has shown that low-metallicity AGNs can overlap with the classical star-forming locus in the N2-BPT diagram (M. Hirschmann et al. 2017, 2019, 2023b), meaning that they may not be removed by the G. Kauffmann et al. (2003) classification.

### 4.2.2 *Relation of strong-line luminosities to SFRs and gas-phase metallicity at low redshift*

In this section, we demonstrate that LUMEN reproduces the H $\alpha$ luminosity function at $z = 0$ and the local scaling relations between strong-line luminosities, SFRs, and gas-phase metallicities. In all panels, we distinguish between simulated galaxies above $10^8\,\mathrm{M}_\odot$ (solid blue median) and those above $10^7\,\mathrm{M}_\odot$ (dotted blue median with shaded area indicating one standard deviation). The top left panel in Fig. 6 shows the H $\alpha$ luminosity function compared to dust-corrected observational results from S. S. Fujita et al. (2003; orange dotted), P. G. Perez-Gonzalez et al. (2003; orange dashed), and C. Ly et al. (2007; orange solid), with exact redshifts indicated in parentheses. The determinations at $z = 0.08$ and $z = 0.026$ differ by more than 1 dex at the faint end, indicating that the low-redshift luminosity function remains poorly constrained. This is likely due to methodological uncertainties, such as cosmic variance, dust corrections, filter profiles, and completeness. The LUMEN prediction lies among the observed luminosity functions, but slightly overpredicts the high-luminosity end at $L_{\mathrm{H}\alpha} > 10^{41.5}$ erg s$^{-1}$.

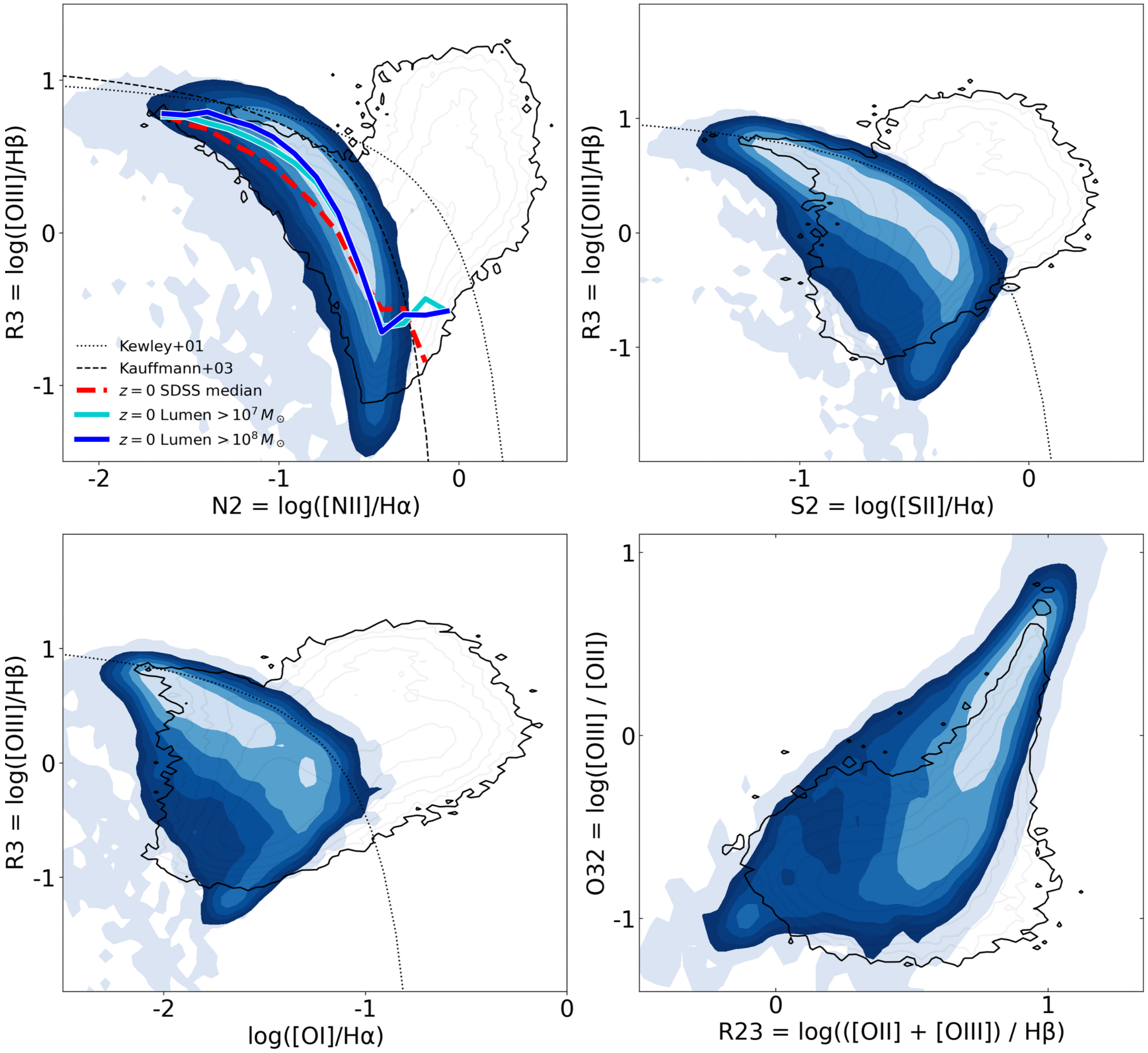

**Figure 5.** Location of LUMEN galaxy populations at $z = 0$ in different line-ratio diagrams, N2-BPT (top left), S2-VO87 (top right), [O I]$\lambda$6300/H$\alpha$-VO87 (bottom left), and O32–R23 (bottom right). Shown are simulated galaxies above a mass cut of $10^8$ M$_\odot$ with line luminosities limited to $10^{39}$ erg s$^{-1}$ (dark blue contours), as well as an extended population down to $10^7$ M$_\odot$ with no luminosity limit (light blue contours). Included for comparison are local SDSS-observed galaxies (black outline with light grey contours), theoretical upper limits for star-forming galaxies (L. J. Kewley et al. 2001a; dotted lines), and the empirical lower limit for AGNs (G. Kauffmann et al. 2003; dashed line). In the N2-BPT diagram, we include median relations for the star-forming branch of the SDSS data (red dashed line), the LUMEN galaxies above $10^8$ M$_\odot$ (dark blue line), and the LUMEN galaxies above $10^7$ M$_\odot$ (light blue line).

We further present the relations between the SFR, the H$\alpha$ luminosity (middle panel), and the combined [O III] + H$\beta$ luminosity ($L_{[\mathrm{O\,III}]+\mathrm{H}\beta}$, bottom panel). Our predictions follow the empirical calibrations from R. C. Kennicutt & N. J. Evans (2012) and D. E. Osterbrock & G. J. Ferland (2006) closely at SFR $> 10^{-2}$ M$_\odot$ yr$^{-1}$, while at lower SFRs the imposed mass cuts cause a mild bias towards higher line luminosities.

The middle and right columns show LUMEN predictions for strong-line metallicity indicators N2, O32, R23, R2 ([O II] $\lambda\lambda$3727, 3729/H$\beta$), R3, and N2O2 ([N II] $\lambda$6584/[O II] $\lambda\lambda$3727, 3729) against various empirical and theoretical calibrations (thin dashed orange, grey, and purple lines; L. J. Kewley & M. A. Dopita 2002; M. Pettini & B. E. Pagel 2004; F. Bresolin 2008; T. Nagao et al. 2006; R. Maiolino et al. 2008; E. Pérez-Montero & T. Contini 2009; T. Jones et al. 2015; M. Curti et al. 2020). Across all diagnostics, LUMEN predictions show good agreement. O32 and R3 exhibit slightly flatter slopes than the observational determinations, but still remain consistent with the 1$\sigma$ scatter. The lowest O/H abundance for the $> 10^8$ M$_\odot$ galaxies lies around $12 + \log(\mathrm{O/H}) \simeq 8.1$, whereas extending the sample to $10^7$ M$_\odot$ adds lower-metallicity systems down to $12 + \log(\mathrm{O/H}) \simeq 7.2$, which continue to follow the literature calibrations closely.

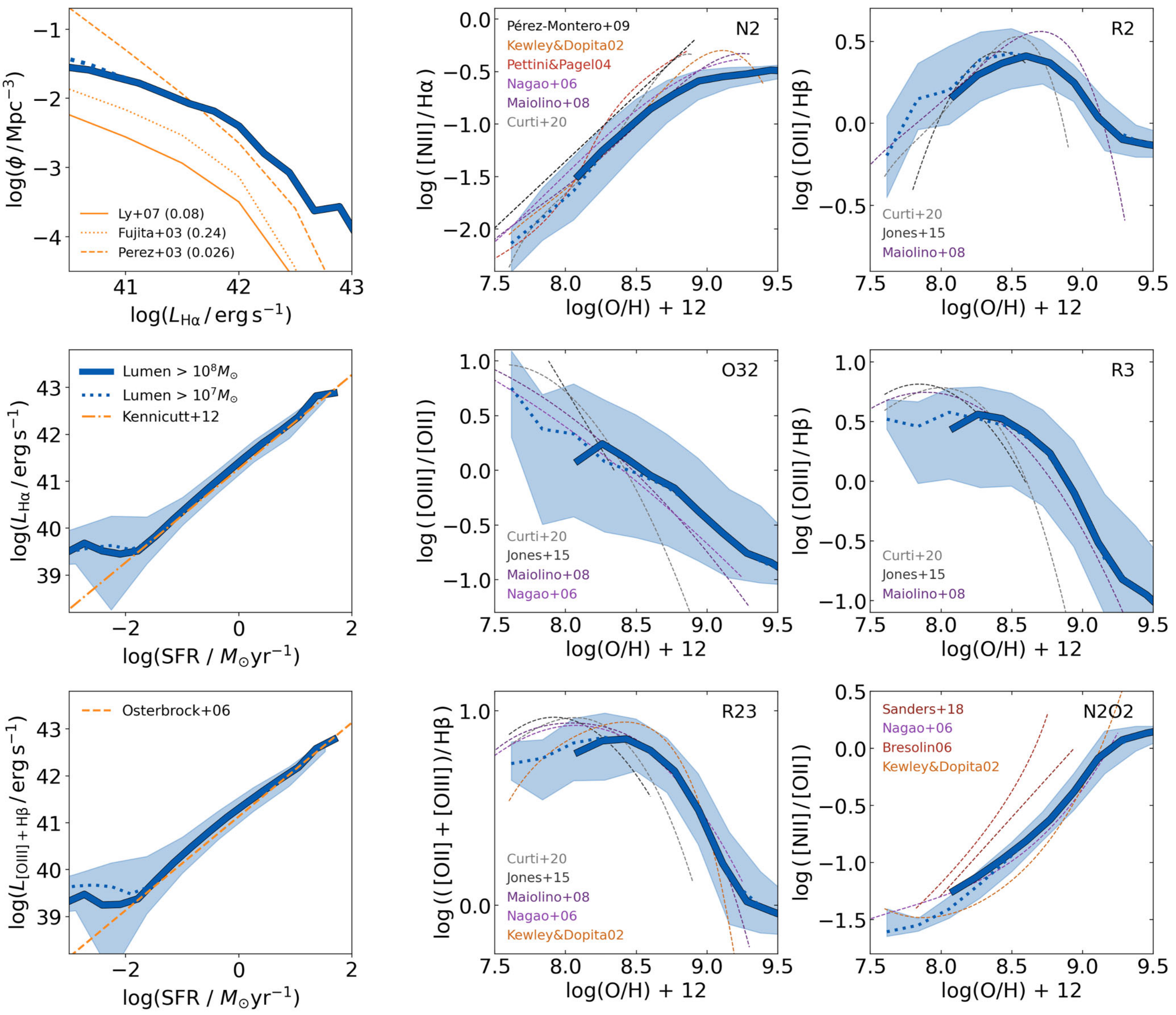

**Figure 6.** Comparison of LUMEN predictions at $z = 0$ against the H $\alpha$ luminosity function (top left), as well as local scaling relations for the SFR (centre and bottom left) and gas-phase metallicity (centre and right columns). Simulated galaxies are shown for mass cuts of $10^8$ $M_\odot$ (solid blue) and $10^7$ $M_\odot$ (dotted blue with 1$\sigma$ scatter). Observational determinations include three locally derived H $\alpha$ luminosity functions (orange lines; S. S. Fujita et al. 2003; P. G. Perez-Gonzalez et al. 2003; C. Ly et al. 2007), and widely used calibrations for the SFR–H $\alpha$ (centre left; R. C. Kennicutt & N. J. Evans 2012) and combined H $\beta$+[O III]–SFR relation (bottom left; D. E. Osterbrock & G. J. Ferland 2006). Metallicity diagnostics N2 (top middle), O32 (centre), R23 (bottom middle), R2 (top right), R3 (centre right), and N2O2 (bottom right) are compared to various local calibrations (thin dashed grey, purple, and orange lines; L. J. Kewley & M. A. Dopita 2002; M. Pettini & B. E. Pagel 2004; F. Bresolin 2008; T. Nagao et al. 2006; R. Maiolino et al. 2008; E. Pérez-Montero & T. Contini 2009; T. Jones, C. Martin & M. C. Cooper 2015; M. Curti et al. 2020).

## 5 RESULTS: HIGH-$Z$ EMISSION-LINE RATIOS IN DIFFERENT LUMEN VARIATIONS

Given the successful validation of the LUMEN Base model at $z = 0$, we extend our predictions to $z = 3$–8. We compare predictions from different LUMEN variants (Section 3.4, Table 2) to recent *JWST* observations to identify which changes in ionizing spectra or gas conditions might drive the observed line ratios in the N2-BPT (Section 5.1), S2-VO87 (Section 5.2), and O32–R23 (Section 5.3) diagrams. The stellar masses and star formation rates of the observational samples fall broadly within the parameter space covered by the TNG50 sample. We treat the observed $z = 3$–8 population as a single sample, since current observations show no significant evolution in emission-line properties across this redshift range (A. E. Shapley et al. 2025; L. Clarke et al. 2026). In TNG50, the main exception is the number of line-emitting AGNs, which increases from fewer than 100 at $z > 4$ to more than 1100 at $z \leq 4$.

For comparison, predictions at $z = 2$ are presented separately in Appendix A. At this redshift, the observed N2-BPT, S2-VO87, and O32–R23 distributions can be reproduced by several partially degenerate combinations of effects, which limits the ability of current data to identify the dominant physical drivers.

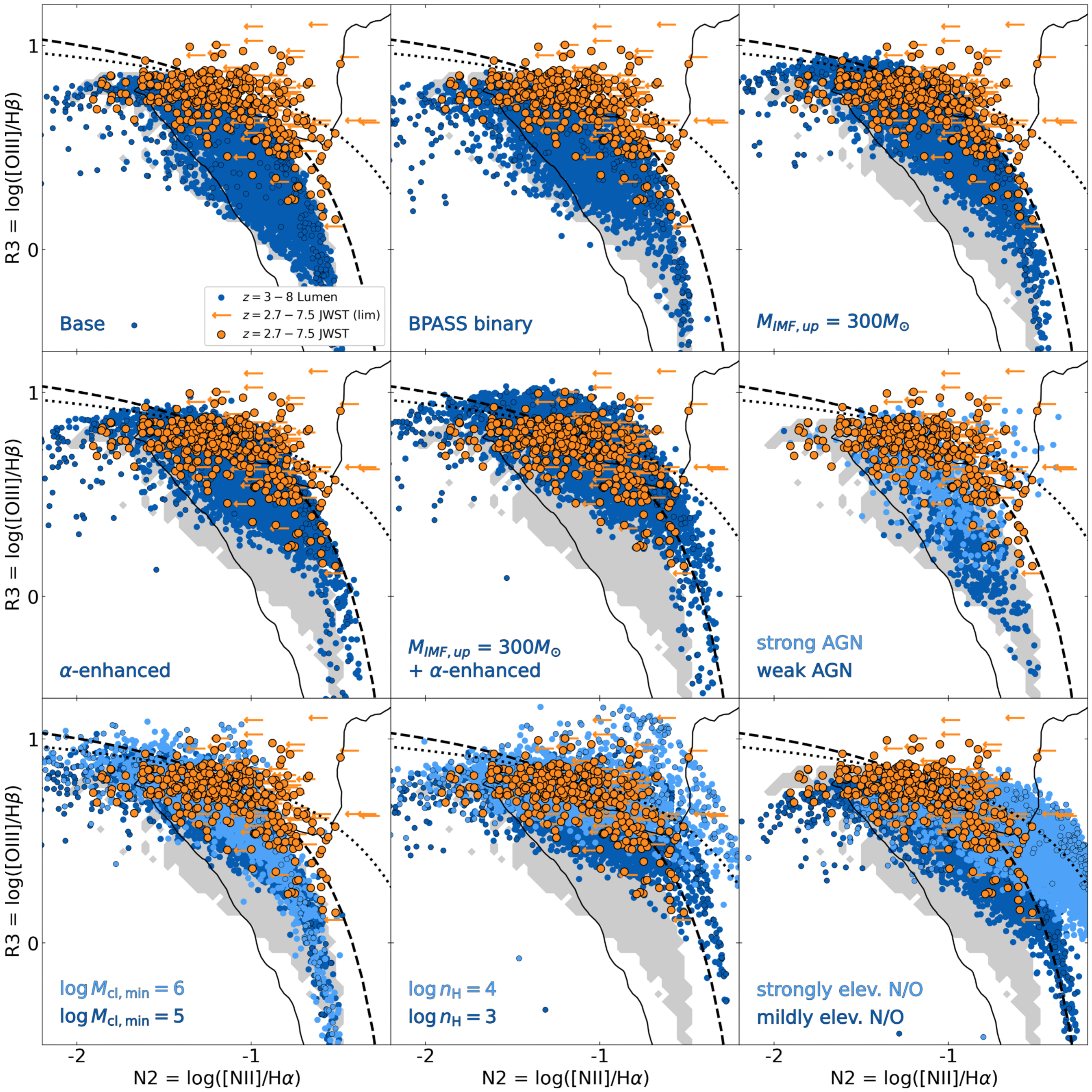

**Figure 7.** Location of galaxy populations at $z = 3$–8 from various LUMEN models (light and dark blue scatter, flux-limited to $10^{-19}$ erg s$^{-1}$ cm$^{-2}$) in the classical N2-BPT diagram. LUMEN variations shown are: Base model (top left), BPASS (top middle), high IMF upper-mass cutoff (top right), stellar $\alpha$-enhancement (middle left), high IMF upper-mass cutoff and stellar $\alpha$-enhancement combined (centre), weak and strong AGNs (middle right), minimum star-cluster masses of $10^5$ and $10^6$ M$_\odot$ (bottom left), $n_{\rm H}$ densities of $10^3$ and $10^4$ cm$^{-3}$ (bottom middle), and mildly and strongly elevated N/O abundances (bottom right). Re-simulated ILLUSTRISTNG100 galaxies are indicated with black edge colours. In order to visualize changes due to the parameter variations, all panels include the Base model predictions (grey contour). *JWST* observations and upper limits (orange) spanning $z = 2.7$–7.5 have been compiled from the JADES (A. J. Cameron et al. 2023; L. Clarke et al. 2026), CEERS (R. L. Sanders et al. 2023a), and AURORA (A. E. Shapley et al. 2025) programs. Included for comparison are SDSS galaxies (black outline) and local ionizing-source classification lines (dashed and dotted lines).

### 5.1 N2-BPT diagram

In Fig. 7, we present the predicted locations of simulated galaxies from various LUMEN model configurations at $z = 3$–8 in the N2-BPT diagram. Each panel isolates a single model variation shown in blue, while in some cases a more extreme version of the same modification is included in lighter blue. Predictions from the Base model are included as a grey-shaded contour in all panels. The 50 ILLUSTRISTNG100 zooms from the BLACKDAWN simulation suite are indicated with a black outline and generally follow the same distribution as the main TNG50 population. From left to right, the top row compares the Base model to variants using BPASS binary stellar populations and an increased IMF upper-mass cutoff of 300 M$_\odot$. The middle row shows stellar $\alpha$-enhancement, its combination with an increased

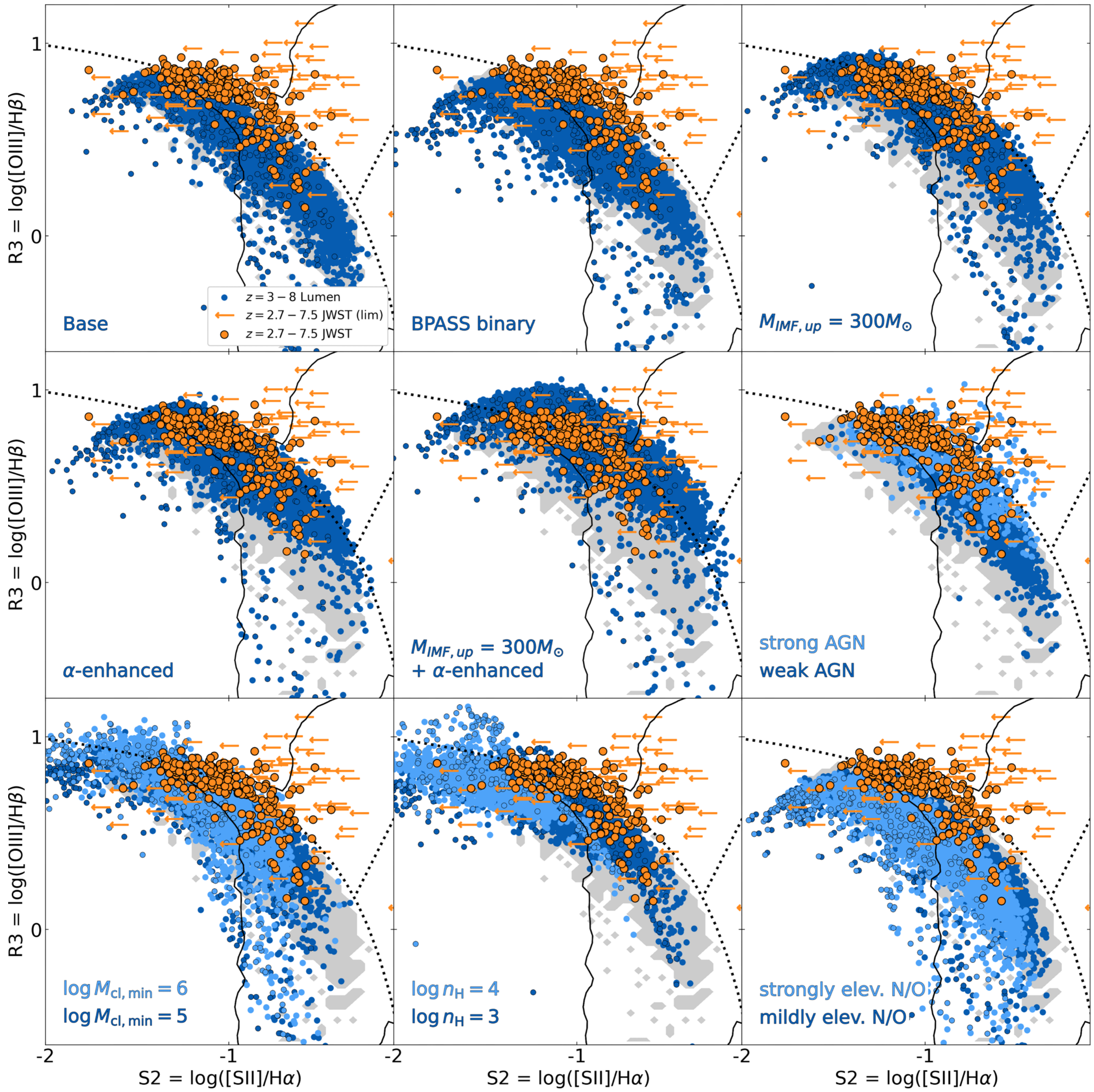

**Figure 8.** Location of galaxy populations at $z = 3$–$8$ from the same LUMEN models as in Fig. 7 (blue data points, flux-limited to $10^{-19}$ erg s$^{-1}$ cm$^{-2}$) in the S2-VO87 diagram, compared against *JWST* observations (orange) between $z = 2.7$ and 7.5 from the JADES, CEERS, and AURORA (A. J. Cameron et al. 2023; R. L. Sanders et al. 2023a; A. E. Shapley et al. 2025; L. Clarke et al. 2026) programs. As before, all panels include the Base model predictions (grey contour), SDSS galaxies (black outline), and local ionizing-source classification boundaries (dotted lines).

IMF upper-mass cutoff, and weak and strong AGN contributions. The bottom row explores increased minimum star-cluster masses of $10^5$ and $10^6$ M$_\odot$, high ionized gas densities of $10^3$ and $10^4$ cm$^{-3}$, and mild and strong nitrogen enrichment. A flux limit of $10^{-19}$ erg s$^{-1}$ cm$^{-2}$ is applied to all emission lines, mimicking an approximate NIRSpec detection limit at these redshifts and wavelengths (A. E. Shapley et al. 2025). We compare to recent *JWST* measurements of star-forming galaxies spanning $z = 2.7$–7.5 from the AURORA, CEERS, and JADES programs (orange; A. J. Cameron et al. 2023; R. L. Sanders et al. 2023a; A. E. Shapley et al. 2025; L. Clarke et al. 2026). As in Fig. 5, SDSS galaxies and standard local classification lines are shown to guide the eye.

In the Base model, galaxies move along the local star-forming branch with increasing redshift, towards higher R3 and lower N2, indicating lower metallicities and higher excitation. However, most *JWST* sources lie above this branch and thus the Base model fails to reproduce the majority of observations. This suggests that the ionizing processes and gaseous environments responsible for the observed high-$z$ line ratios differ substantially from local conditions. The BPASS binary model performs very similarly to the

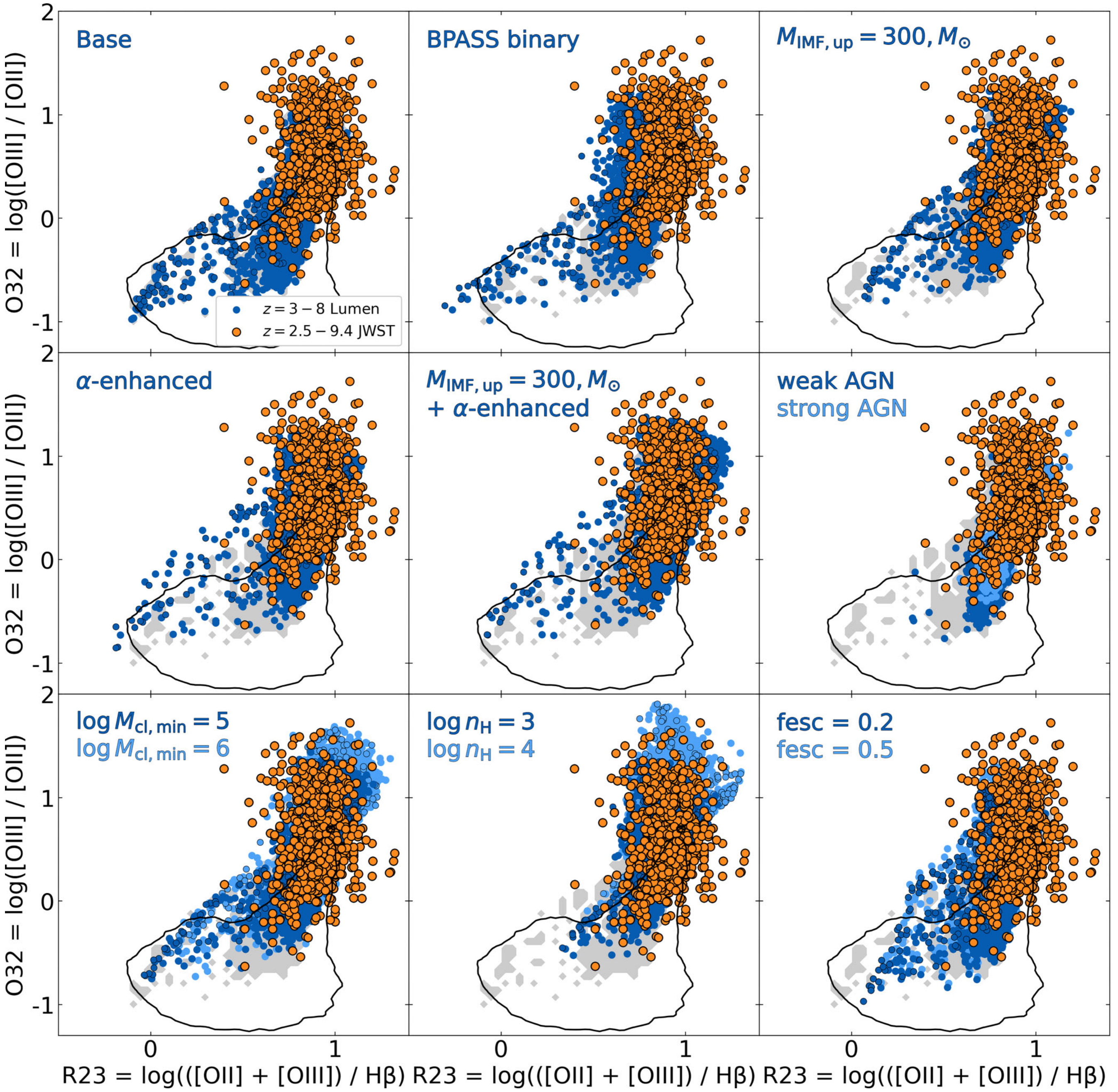

**Figure 9.** Location of LUMEN galaxy populations at $z = 3$–8 (blue data points, flux-limited to $10^{-19}$ erg s$^{-1}$ cm$^{-2}$) in the O32–R23 diagram. Shown are the Base model (top left), BPASS (top middle), high IMF upper-mass cutoff (top right), stellar $\alpha$-enhancement (middle left), high IMF upper-mass cutoff and stellar $\alpha$-enhancement combined (centre), weak and strong AGNs (middle right), increased minimum cluster masses to $10^5$ and $10^6$ M$_\odot$ (bottom left), increased $n_{\rm H}$ density to $10^3$ and $10^4$ cm$^{-3}$ (bottom middle), and density-bounded H II regions with escape fractions of 0.2 and 0.5. The compilation of high-$z$ observations includes various *JWST* data sets between $z = 2.7$ and 9.4 (orange; M. Curti et al. 2022; A. J. Cameron et al. 2023; S. Mascia et al. 2023; R. L. Sanders et al. 2023a; K. Boyett et al. 2024; G. Roberts-Borsani et al. 2024; A. Calabrò et al. 2024a, b; A. E. Shapley et al. 2025; L. Clarke et al. 2026). For reference, the local SDSS data (black outline) and Base model predictions (grey contour) are displayed in the background.

Base model, which relies on single-star SEDs from G. Bruzual & S. Charlot (2003). As discussed in Section 3.4, this is because optical lines below the He II ionizing threshold are relatively insensitive to the inclusion of harder ionizing radiation from binary stars. Additional differences in the Wolf–Rayet star modelling between BPASS and the G. Bruzual & S. Charlot (2003) SEDs further compensate most of the binary-star impact in this diagram.

Using an IMF upper-mass cutoff of 300 M$_\odot$ produces a systematic increase in R3 due to harder ionizing radiation from more massive stars. This shifts our predictions towards the AGN boundary, now matching observations with moderate N2 ($< -0.9$) and high R3 ($\sim 0.8$), but not more extreme values. While harder ionizing radiation via stellar $\alpha$-enhancement is frequently inferred as the main driver of elevated R3 at $z \gtrsim 2$ (C. C. Steidel et al. 2016; A. L. Strom et al. 2017; A. E. Shapley et al. 2019, 2025;

R. L. Sanders et al. 2020, 2023a; M. W. Topping et al. 2020; J. N. Runco et al. 2021), we only find moderate N2 and R3 increases, similar to the model with a high IMF upper-mass cutoff. We note again that the employed level of stellar $\alpha$-enhancement is relatively extreme at $Z_\star = Z_{\rm gas}/5$ and therefore already represents a limiting case. Only combining stellar $\alpha$-enhancement with a higher $m_{\rm up}$ reproduces the concentration of observed galaxies around R3 $\sim 0.8$ and reaches several of the higher N2 values and upper limits near R3 $\sim 1$. However, even this extreme hardening of the ionizing spectrum fails to reach the highest observed R3 and N2 values.

Galaxies with a weak AGN contribution largely overlap with the Base model, as line strengths are only marginally affected. In contrast, strong AGN contributions produce sources that extend beyond the classical AGN boundary and match part of the observed high-$z$ scatter. Their low metallicities shift them to lower N2 compared to local SDSS AGNs, which places them closer to high-$z$ star-forming galaxies (Y. Harikane et al. 2023; M. Hirschmann et al. 2023b). Unidentified AGN contamination can therefore produce line ratios similar to the *JWST* observations. However, the simulated galaxies classified as having strong AGN contributions are found predominantly in the $z = 3$–4 snapshots. While weaker AGN contributions are still present at higher redshifts, these do not significantly alter the predicted line ratios. Strong AGN contamination may therefore explain some of the observed scatter at $z \leq 4$, but is unlikely to account for the more extreme observations at $z > 4$. In addition, it remains uncertain whether TNG50 accurately predicts the abundance of high-$z$ AGNs.

Increasing the minimum star-cluster mass in LUMEN boosts $\log U$ by concentrating stellar mass into fewer, more massive clusters. At fixed SFR, this implies fewer H II regions, but with higher ionizing photon flux per region. Setting the minimum star-cluster mass to $10^5$ M$_\odot$, or even $10^6$ M$_\odot$, shifts the entire population towards higher R3, with some values exceeding 1. However, this version does not account for high N2 values ($> -0.9$).

The LUMEN model with elevated ionized gas density of $n_{\rm H} = 10^3$ cm$^{-3}$ reproduces moderate R3 and N2 at a level comparable to the model with combined hardening from high $m_{\rm up}$ and stellar $\alpha$-enhancement. Increasing the density further to $n_{\rm H} = 10^4$ cm$^{-3}$ matches the most extreme R3 and N2 values of the observed distribution. This is because at fixed volume-filling factor, increasing the density increases $\log U$, which boosts R3. Since both $10^3$ and $10^4$ cm$^{-3}$ remain below the [N II]$\lambda$6584 critical density ($\sim 10^5$ cm$^{-3}$), N2 also increases with density. However, such high densities decrease the S2 ratio, in tension with $z \geq 3$ observations in the S2-VO87 diagram (Fig. 8). A uniformly high ionized-gas density therefore cannot simultaneously reproduce the observed N2-BPT and S2-VO87 distributions.

N2 can also be increased via elevated N/O at fixed O/H. Both mildly and strongly enhanced models shift LUMEN predictions to higher N2 at fixed moderate R3 ($\sim 0.5$–0.8), thereby successfully matching high-N2 systems. Only the strongly enhanced model, with a 0.6 dex increase in N/O, reaches the highest N2 values around $-0.5$. This requires $\log({\rm N/O}) \sim -0.5$ to 0.5 at $12 + \log({\rm O/H}) \geq 8$, consistent with observed high-$z$ N/O abundances (R. Marques-Chaves et al. 2024; E. Cataldi et al. 2026; I. Morel et al. 2026). However, neither model reproduces galaxies with both high R3 and low N2.

In summary, no single mechanism reproduces the full observed distribution at $z > 3$. The Base and BPASS models fail almost entirely, while harder ionizing spectra (from a higher IMF upper-mass cutoff and/or stellar $\alpha$-enhancement) can produce moderately elevated R3 and N2. Weak AGNs do not change the overall line contributions significantly, but if not properly excluded from the sample, strong AGN contributions may account for some of the more extreme systems at $z \leq 4$. Reaching the highest R3 values ($> 1$) likely requires elevated ionization parameters in addition to harder spectra, potentially driven by systematically more massive star clusters at high redshift, which have been observed with *JWST* (E. Vanzella et al. 2023; A. Adamo et al. 2024; M. Messa et al. 2024; L. Mowla et al. 2024; A. Claeyssens et al. 2025, 2026). The highest N2 values ($> -0.9$) require either significant nitrogen enrichment or increased gas densities. Although $n_{\rm H} = 10^4$ cm$^{-3}$ can reproduce both high R3 and N2, it suppresses S2, in tension with the observed S2 distribution (Fig. 8).

### 5.2 S2-VO87 diagram

Fig. 8 shows the S2-VO87 diagram for the same model configurations as in Fig. 7. It compares LUMEN predictions at $z = 3$–8 to *JWST* observations, SDSS galaxies, and standard classification lines. As in the N2-BPT diagram, the Base and BPASS models reproduce only galaxies below the starburst line (L. J. Kewley et al. 2001a), matching moderate R3 and S2 but failing to capture the observed scatter above it. Models with a higher IMF upper-mass cutoff or stellar $\alpha$-enhancement partially reproduce systems just above the AGN line (moderate S2 $< -0.8$, R3 $\sim 0.5$–0.8), but do not reach the more strongly offset sources. We caution that our $\alpha$-enhanced models do not self-consistently alter the gas-phase abundance pattern. Since part of sulphur production is associated with Type Ia supernova enrichment (C. Kobayashi et al. 2020), truly $\alpha$-enhanced systems may be sulphur-deficient relative to oxygen. This could lower S2 compared to our predictions, consistent with recent evidence for sulphur deficiency at cosmic noon (N. S. J. Rogers et al. 2026).

Combining a higher IMF upper-mass cutoff and stellar $\alpha$-enhancement yields a good match across the R3–S2 plane, where only the most extreme case (R3 $\sim 1.2$ with S2 upper limit) remains unexplained. Weak AGNs show minimal impact, while strong AGNs shift many galaxies beyond the starburst line and may account for part of the observed scatter. As for the N2-BPT diagram, almost all of these galaxies come from the $z = 3$ and 4 snapshots and therefore cannot explain the $z > 4$ observations.

Increasing the ionization parameter through massive star clusters ($> 10^5$–$10^6$ M$_\odot$) again produces R3 $\gtrsim 1$, which might explain some high-R3 systems. However, S2 is slightly reduced, as higher $\log U$ shifts sulphur to higher ionization states. This prevents the model from matching most S2 observations beyond the starburst line. Models with increased ionized gas densities also reduce S2. This effect is strongest at $10^4$ cm$^{-3}$, well above the [S II]$\lambda\lambda$6717, 6731 critical density ($\sim 10^3$ cm$^{-3}$), where collisional de-excitation suppresses emission. At $10^3$ cm$^{-3}$, near the critical density, this suppression is weaker. As reproducing *JWST* observations requires elevated S2 relative to local measurements, this rules out uniformly high-density scenarios. LUMEN's nitrogen-enriched models also exhibit a modest S2 decrease. This mainly reflects the global rescaling of other elemental abundances to maintain constant total metallicity at elevated nitrogen.

Overall, we conclude that reproducing the full observed range in the S2-VO87 diagram requires harder ionizing radiation due to both a higher IMF upper-mass cutoff and stellar $\alpha$-enhancement, potentially combined with elevated ionization parameters through high star-cluster masses, or with strong AGN contributions.

### 5.3 The O32–R23 diagram

Lastly, we turn to the O32–R23 diagram. Fig. 9 shows the predicted locations of simulated galaxies in the O32–R23 diagram at $z = 3$–8. As in Figs 7 and 8, fluxes are limited to $10^{-19}\,\mathrm{erg\,s^{-1}\,cm^{-2}}$. The panel layout follows the same model variations, except for the bottom right, which assumes density-bounded H II regions with LyC escape fractions of 0.2 and 0.5. We compare to recent *JWST* observations spanning $z \simeq 2.7$–9.4 (orange; M. Curti et al. 2022; A. J. Cameron et al. 2023; S. Mascia et al. 2023; R. L. Sanders et al. 2023a; K. Boyett et al. 2024; G. Roberts-Borsani et al. 2024; A. Calabrò et al. 2024a, b; A. E. Shapley et al. 2025; L. Clarke et al. 2026), with the local SDSS distribution shown for reference (black outline).

At $z = 3$–8, the Base-model galaxies shift towards higher O32 and R23 values compared to $z = 0$, consistent with the transition to lower metallicity and higher ionization conditions at early epochs. The highest predicted O32 values in the Base model remain around $\sim 1$, similar to the upper end of the $z = 0$ distribution, which indicates an absence of very high-ionization systems. The BPASS variant slightly broadens the distribution towards higher O32 and lower R23, now reaching O32 $\sim 1.2$. Increasing the IMF upper-mass cutoff or accounting for stellar $\alpha$-enhancement similarly extends O32 to $\sim 1.2$. Combining both effects increases O32 to $\sim 1.3$, but the predictions remain far below the observed values out to $\sim 1.8$. This demonstrates that harder ionizing spectra alone cannot reproduce the full observed distribution. Introducing weak and strong AGN components does not change the predicted line ratios significantly, with only a single source in the strong AGN model producing an elevated O32 value ($\sim 1.2$). While this indicates that AGNs can elevate O32 values, we do not predict AGNs to account for a significant fraction of the observed high-O32 systems.

As O32 traces the ionization parameter $\log U$, increasing the minimum star-cluster mass strongly boosts O32 values. Raising the mass threshold from $10^3$ to $10^5$ and $10^6\,\mathrm{M_\odot}$ shifts the galaxy population towards high O32, with the $10^6\,\mathrm{M_\odot}$ model reaching up to $\sim 1.7$, in good agreement with the observed sample. The model with an elevated ionized gas density of $10^4\,\mathrm{cm^{-3}}$ also reaches the most extreme O32 values, with some values extending even beyond the observed range. This stems from the dual effect of increasing ionization parameters and suppressing the combined [O II]$\lambda\lambda3727, 3729$ strength due to the low critical density of [O II]$\lambda3729$ ($3 \times 10^3\,\mathrm{cm^{-3}}$; D. W. Hogg et al. 1998). Another consequence of the decreased [O II]$\lambda\lambda3727, 3729$ doublet is that the model locus is shifted towards lower R23, which misses the observed high R23 ($\sim 1$) at moderate O32 ($< 1$). The density adopted in the $n_\mathrm{H} = 10^3\,\mathrm{cm^{-3}}$ model remains below the critical density of [O II]$\lambda3729$ and as a result, O32 is only slightly increased to 1.3, owing to the higher ionization parameter.

Both LUMEN models with a non-zero escape fraction produce only a modest increase in O32 relative to the Base model. This is because, at lower ionization parameters ($\log U < -2$), O32 does not increase significantly in the density-bounded H II-region models, as discussed in A. Plat et al. (2019). Although individual H II regions with $\log U \geq -2$ reach higher O32, the integrated galaxy emission is dominated by the larger population of lower-$\log U$ regions. LyC leakage alone, without systematically elevated ionization parameters, therefore cannot explain the O32 observations at $z \geq 3$.

In summary, harder radiation fields due to stellar $\alpha$-enhancement, a high IMF upper-mass cutoff, and potentially AGNs can increase O32, but achieving the full dynamic range observed by *JWST* requires elevated ionization parameters, likely produced by more massive star clusters at high redshift. High ionized gas densities in the H II regions may explain the most extreme O32 values, but only if the observed objects do not simultaneously show moderate to high S2 values, as these are suppressed at $n_\mathrm{H} \geq 10^3\,\mathrm{cm^{-3}}$ (see Section 5.2).

### 5.4 The 'Master' model for high-$z$

As shown in the previous sections, no single parameter variation can reproduce the full range of observed high-$z$ emission-line ratios in all three diagrams. Different subsets of the observed population are best reproduced by different physical mechanisms, some pointing to harder ionizing spectra, others to higher ionization parameters via increased star-cluster masses, and still others to elevated N/O abundance. To explore whether a combined model can simultaneously predict all observed trends, we constructed a composite 'Master' model. This model incorporates the key ingredients identified as likely necessary to reproduce the full range of observed line ratios in the previous analyses: a higher IMF upper-mass cutoff ($300\,\mathrm{M_\odot}$), stellar $\alpha$-enhancement ($Z_\star = Z_\mathrm{gas}/5$), increased ionization parameters through massive star clusters ($M_\mathrm{cl,min} = 10^5\,\mathrm{M_\odot}$), and elevated nitrogen abundances (+0.1 dex N/O at fixed O/H).

The resulting model represents an intentionally extreme case. Applying all parameter changes uniformly to every galaxy is not meant to describe a realistic population, but rather to illustrate how far the combined impact of several physical processes could offset galaxies from the $z = 0$ Base model. The actual high-$z$ population likely spans a continuum of properties between the Base model and the extreme variations encompassed by the 'Master' configuration. Accurately recovering this distribution would require simulations that self-consistently track the evolution and interplay of these mechanisms, which current models are not yet capable of capturing.

However, several of the relevant processes are likely correlated. Massive and compact star clusters are expected to form preferentially in turbulent, dense, high-pressure ISM conditions, as found at $z > 0.7$ and high SFR surface density (A. Claeyssens et al. 2025). N. A. Reddy et al. (2023) further demonstrated a positive correlation between star formation rate surface density, $\log U$, and $n_\mathrm{H}$ at $z = 2.7$–6.3, consistent with the interpretation that compact, intensely star-forming regions can drive high ionization parameters. The most massive star clusters may naturally host more massive stars, either through stochastic sampling of the IMF or through a possible correlation between the maximum stellar mass and the total star-cluster mass (C. Weidner & P. Kroupa 2006). In the densest systems, mass segregation and stellar collisions could further promote the formation of very massive and supermassive stars (M. Gieles et al. 2018), which may enhance nitrogen production in the most compact regions (R. Marques-Chaves et al. 2024). Future model explorations might therefore couple the gas density, star-cluster mass, IMF upper-mass cutoff, and N/O abundance, rather than imposing the same extreme parameter combination on the entire galaxy population.

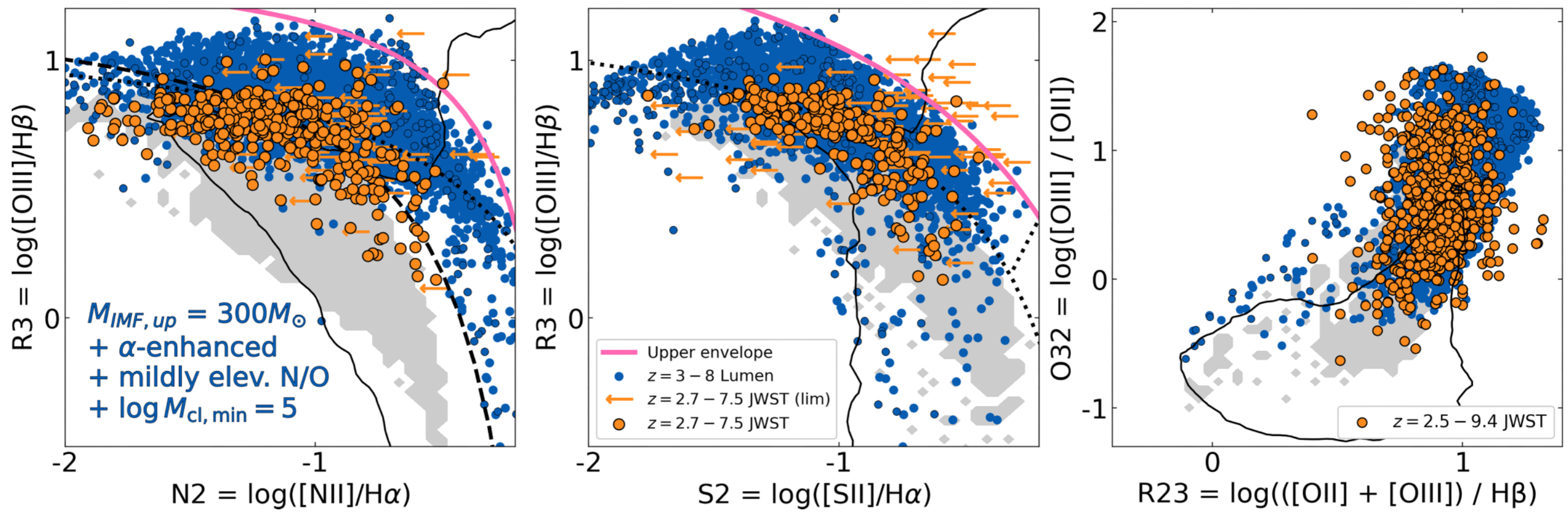

**Figure 10.** Location of galaxy populations at $z = 3$–8 from LUMEN Base and 'Master' models (grey contours and blue data points, respectively, flux-limited to $10^{-19}\,\mathrm{erg\,s^{-1}\,cm^{-2}}$) in the N2-BPT (left), S2-VO87 (middle), and O32–R23 (right) diagrams against observations between $z = 2.5$and 9.4 (orange). In addition to the classical classification lines from L. J. Kewley et al. (2001b) and G. Kauffmann et al. (2003), new demarcation lines are shown for the N2-BPT and S2-VO87 diagrams (pink).

#### 5.4.1 *Demarcation lines for line-ratio diagrams*

Even with the limitations discussed above, the 'Master' model provides a useful framework to examine how the combined mechanisms manifest in observable line ratios. Fig. 10 compares the predictions from the Base (grey contours) and 'Master' (blue scatter) models in the N2-BPT, S2-VO87, and O32–R23 diagrams at $z = 3$–8. In contrast to the previously shown LUMEN variations, the 'Master' model simultaneously reaches the most extreme R3, N2, S2, and O32 ratios of the $z \geq 3$ observations. The model populations are shifted somewhat beyond the observed loci in all three diagrams, indicating that these most extreme parameter variations indeed should not apply uniformly to all observed galaxies. Most systems likely require weaker or different combinations of these effects.

To define an upper boundary for the location of high-$z$ star-forming galaxies in the N2-BPT and S2-VO87 diagrams, we derive demarcation lines tracing the envelope of the 'Master' model predictions, shown as pink lines in Fig. 10. To determine these lines, we divided the horizontal axis into 25 bins, calculated the 99th percentile of the model distribution in each bin, and fit these points with a function of the form $y = a/(x - b) + c$. Because each line is fitted to the 99th percentile in each bin using a smooth analytical function, a small fraction of model predictions lies above the resulting demarcation line. This approach provides a robust and stable upper boundary that is not influenced by stochastic outliers while still encompassing the vast majority of the star-forming model predictions.

$$\log\left(\frac{[\mathrm{O\,III}]\lambda5007}{\mathrm{H}\beta}\right) = \frac{0.32}{\log\left(\frac{[\mathrm{N\,II}]\lambda6584}{\mathrm{H}\alpha}\right) - 0.10} + 1.43 \quad (3)$$

$$\log\left(\frac{[\mathrm{O\,III}]\lambda5007}{\mathrm{H}\beta}\right) = \frac{1.63}{\log\left(\frac{[\mathrm{S\,II}]\lambda\lambda6717,6731}{\mathrm{H}\alpha}\right) - 0.86} + 1.93. \quad (4)$$

#### 5.4.2 *Evolution of Hα and [O III]λ5007 line luminosity functions*

To further test our 'Master' model, we show the predicted evolution of the H$\alpha$ (top row) and [O III]λ5007 (bottom row) luminosity functions (LFs) between $z = 3$ and 8 for both the Base (dark blue) and 'Master' (light blue) LUMEN models in Fig. 11, compared with dust-corrected observational estimates (R. A. Meyer et al. 2024; A. Covelo-Paz et al. 2025; R. A. Meyer et al. 2026), and an extrapolation from the UV luminosity function (S. De Barros et al. 2019).

Both models reproduce the general evolution of the observed luminosity functions across redshift, with the overall normalization and slopes in good agreement where observational determinations are available. For the H$\alpha$ luminosity function, the two models produce very similar predictions. They match the fits from A. Covelo-Paz et al. (2025) particularly well at $z = 5$ and 6, with a slight overprediction at $z = 3$. The difference between the two models is more pronounced for [O III]λ5007 than for H$\alpha$, as the [O III]λ5007 emission is more sensitive to the increased ionization parameter driven by the more massive star clusters in the 'Master' model. At $z = 3$ and 5, the Base model provides a closer match to the observed [O III]λ5007 luminosity function, while at $z = 6$–7 both models reproduce the available data well. By $z = 8$, the 'Master' model agrees better.

We emphasize again that the 'Master' model was constructed to reproduce the current high-$z$ sample of observed optical line ratios. These constraints likely probe only the most luminous end of the emission-line galaxy population. Deeper spectroscopic coverage will be required to determine whether the true population distribution lies closer to the Base, 'Master', or intermediate scenarios.

## 6 DISCUSSION

In Section 5, we showed that different LUMEN model variations reproduce extreme emission-line ratios observed at $z = 2.7$–7.5 to varying degrees. We now interpret these results in a broader physical context. We first examine the origin of the high ionization parameters implied by the observations (Section 6.1), then assess key modelling caveats (Section 6.2), and finally place the LUMEN results in the context of previous theoretical work (Section 6.3).

### 6.1 Increased star-cluster masses at high-$z$ as the origin of high ionization parameters

In Figs 7, 8, and 9, we showed that the most extreme observed R3 and O32 values are reached only in models where the ionization parameter is increased with respect to the Base model for $z = 0$. In this section, we examine this link more directly by analysing how the ionization parameter varies across the different model variants and how this variation imprints on the observed high-$z$ line ratios.

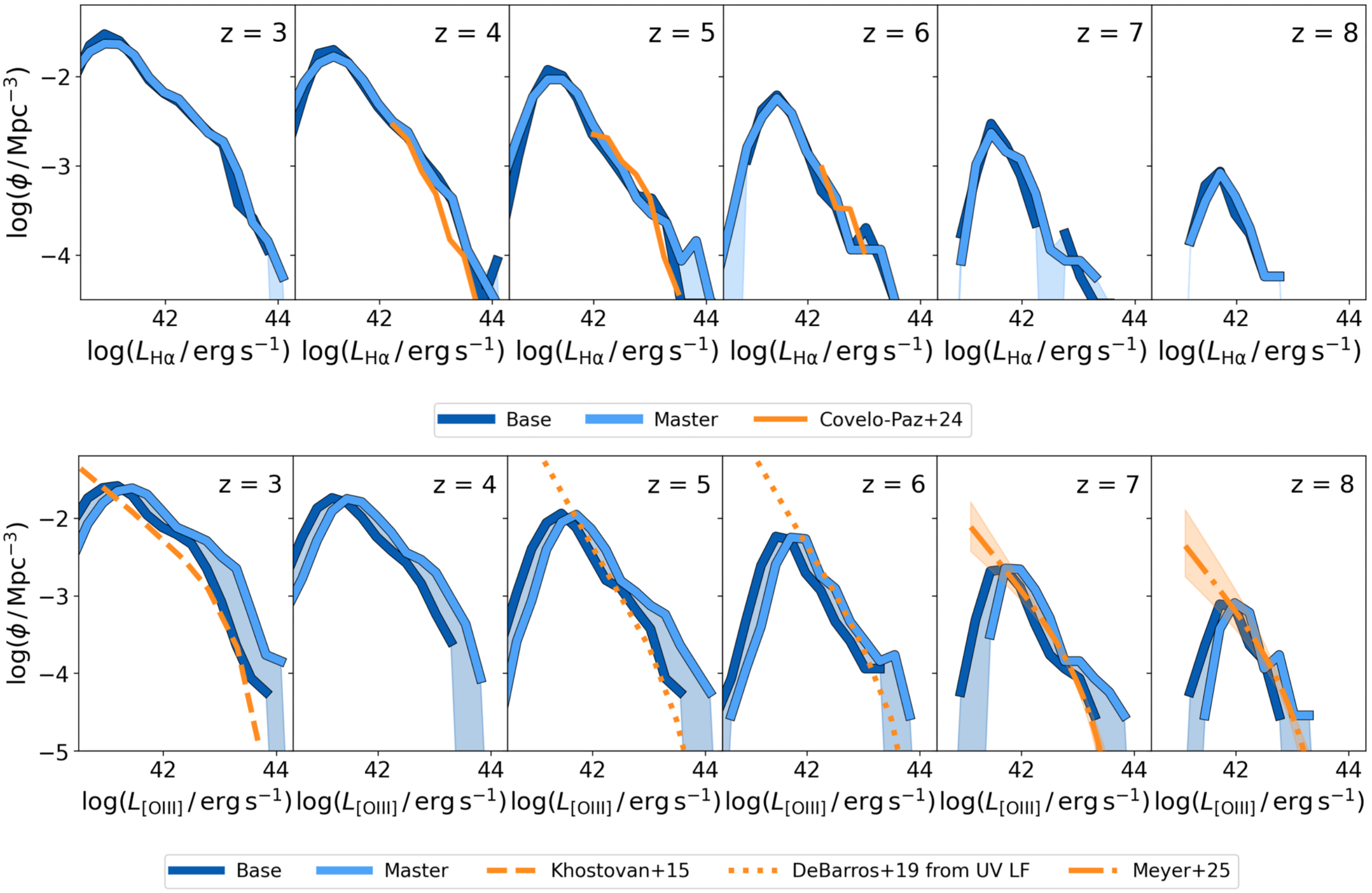

**Figure 11.** Luminosity functions from $z = 3$ to 8 for H $\alpha$ (top row) and [O III]$\lambda$5007 (bottom row), as predicted by the Base (dark blue) and 'Master' (light blue) LUMEN models, compared to dust-corrected observational constraints (orange): H $\alpha$ from A. Covelo-Paz et al. (2025; solid); [O III]$\lambda$5007 from A. A. Khostovan et al. (2015; dashed), and R. A. Meyer et al. (2024, 2026; dash–dotted), respectively, with an additional extrapolation from the UV luminosity function by S. De Barros et al. (2019; dotted).

In Fig. 12, we summarize the relationship between the ionization parameter (top panel) and the R3 (middle panel) and O32 (bottom panel) distributions across the LUMEN model variants at $z = 3$–8. Elevated N/O models are omitted because they do not materially affect R3 or O32. Boxplots show the predicted R3 and O32 values, while orange horizontal lines indicate extreme line ratios from recent *JWST* data sets. Thick solid lines mark the 90th percentile of the observations, while thin lines signify values above this threshold. Blue markers connected with a solid line show the mean $\log U$ with $1\sigma$ scatter in each model, while the dashed line shows the maximum $\log U$.

Fig. 12 confirms that only models that increase the ionization parameter can reproduce the most extreme O32 and R3 values. Increasing the hardness of the ionizing radiation alone is insufficient. In the Base and the BPASS binary models, R3 and O32 values remain close to the 90th-percentile thresholds. The individual and combined high-$m_{\rm up}$ and $\alpha$-enhanced variants, as well as strong AGN contributions, are able to reproduce more extreme R3 values, but still fail to reach R3 $\sim 1.1$ and O32 $> 1.4$. The LyC-leaking models with $f_{\rm esc} = 0.2$ and 0.5 lower R3 relative to the Base model and increase O32 only marginally, again remaining close to the 90th- percentile constraints. Across these models, the mean $\log U$ stays around $-2.5$, with the upper part of the scatter extending to $\log U \sim -2$.

By contrast, models which increase the ionization parameter shift the entire population to higher R3 and O32 values. Imposing a minimum star-cluster mass of $10^5$ $\mathrm{M}_\odot$ increases the predicted ratios to R3 $\sim 1.05$ and O32 $\sim 1.4$. Raising the minimum star-cluster mass further to $10^6$ $\mathrm{M}_\odot$ can produce the most extreme R3 ratios, yet still falls short of the highest O32 measurements. While the $n_{\rm H} = 10^3$ cm$^{-3}$ model reaches R3 $\sim 0.95$ and O32 $\sim 1.3$ at most, increasing the ionized gas density to $10^4$ cm$^{-3}$ enables the models to reproduce essentially the full observed range of R3 and O32. However, as seen in Section 5.2, S2 is suppressed at these densities, indicating that a single ionized gas density component of $n_{\rm H} = 10^4$ cm$^{-3}$ alone cannot simultaneously reproduce the high-$z$ constraints in the N2-BPT and S2-VO87 diagrams. It may still account for individual systems with extreme R3 and O32. The combined 'Master' model, which incorporates increased star-cluster masses, harder ionizing spectra, and elevated N/O abundances, also covers all observed R3 values and misses only the most extreme O32 values near O32 $\sim 1.7$.

Overall, the ionization parameter shifts compared to the Base model are modest, about 0.4–0.6 dex on average. Mean $\log U$ is approximately $-2.1$ in the variants with minimum cluster masses of $10^5$ $\mathrm{M}_\odot$ or ionized gas densities of $n_{\rm H} = 10^3$ cm$^{-3}$, and rises to $\sim -2$ for minimum cluster masses of $10^6$ $\mathrm{M}_\odot$ and $n_{\rm H} = 10^4$ cm$^{-3}$.

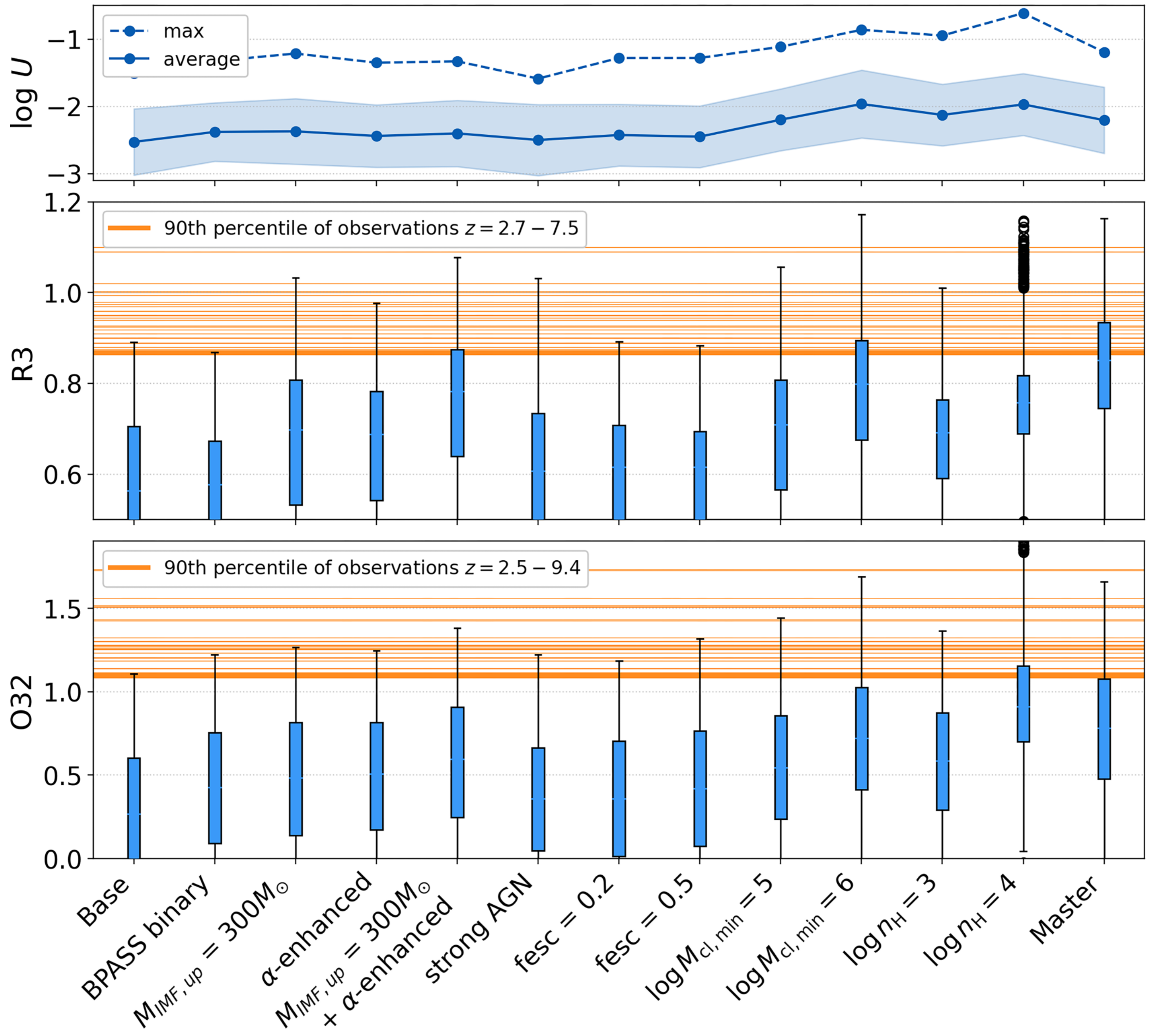

**Figure 12.** R3 (middle panel) and O32 (bottom panel) value distributions for LUMEN models at $z = 3$–$8$ (boxplots, outliers in black circles) compared to the highest observed R3 values at $z = 2.7$–$7.5$ and O32 values at $z = 2.5$–$9.4$ (horizontal orange lines). Shown are the 90th percentiles (thick lines) and values above this threshold (thin lines). In the top panel, the maximum (dashed line) and average (solid line) $\log U$ values of each model are shown, alongside a $1\sigma$ scatter (shaded area).

Owing to its harder ionizing spectrum, the 'Master' model reproduces most observed line ratios at a mean $\log U \sim -2.1$. The highest R3 and O32 values in each model variant arise from the upper tail of the $\log U$ distribution, which reaches approximately $-1$ to $-0.8$, consistent with the most extreme ionization parameters reported for high-$z$ galaxies (e.g. A. J. Cameron et al. 2023; J. A. Zavala et al. 2024; A. Calabrò et al. 2024b; Y. W. Ren et al. 2025).

## 6.2 Caveats

### 6.2.1 *Observational caveats*

Interpreting emission-line ratios at high redshift is affected by several observational limitations. Spectroscopic samples are typically biased towards strong-line emitters, preferentially selecting galaxies with intrinsically extreme ionization conditions. As a result, the observed objects may not be representative of the overall galaxy population. In addition, slit losses and aperture effects mean that observations may probe only part of a galaxy's emission and therefore may not capture its global properties. Imperfect dust corrections can also introduce systematic offsets. However, these caveats do not imply that the observed evolution away from the $z = 0$ and $z \sim 2$ baselines is driven only by systematics, since the most extreme high-$z$ objects occupy a parameter space that is not comparably populated at lower redshift. Furthermore, this work focuses on demonstrating how different physical processes may shift galaxy populations in diagnostic line-ratio diagrams and how extreme line ratios can be produced through a combination of physically plausible mechanisms. Observational uncertainties may affect the inferred population distribution, but are unlikely to erase the qualitative separation between the most extreme high-$z$ systems and the local comparison samples.

### 6.2.2 *Caveats related to TNG50*

TNG50 reproduces a broad set of galaxy population trends across cosmic time, including the star-forming main sequence, galaxy sizes and disc heights, gas fractions and their evolution, as well as rotationally supported kinematics (A. Pillepich et al. 2019; D. Nelson et al. 2019a). Agreement is generally good out to $z \sim 4$–5, but uncertain at higher redshifts due to limited constraints in this regime. Like all current large-volume simulations, TNG50 approximates many physical processes through simplified subgrid prescriptions. Star formation and AGN feedback in particular influence key galaxy properties like star-formation histories, mass–metallicity relations, and gas densities (D. Nelson et al. 2019b). The simulation also does not self-consistently produce enhancements in $\alpha$-element or N/O abundance ratios, so these abundance patterns must be imposed in post-processing through the photoionization models. Capturing such enrichment self-consistently would require revised nucleosynthetic yields from massive stars and Wolf–Rayet phases, improved metal mixing prescriptions, and potentially a more detailed treatment of how feedback redistributes different elements within the ISM.

A further limitation arises from the finite volume of the TNG50 simulation. Observational studies preferentially detect massive, high-sSFR, high-luminosity objects, which are sparsely sampled at the highest redshifts in TNG50. To improve coverage of the high-mass regime, we include 50 of the most massive galaxies from ILLUSTRISTNG100, resimulated as zooms at TNG50 resolution. While many of these systems populate the more extreme end of the predicted emission-line distributions, their intrinsic properties are not sufficiently distinct to separate them clearly from the main TNG50 sample. We therefore do not expect limited volume sampling to significantly bias the trends discussed here.

### 6.2.3 *Caveats related to the photoionization modelling*

In our CLOUDY photoionization models, we further assume idealized 1D spherical H II regions, created by a uniform radiation field incident on constant-density gas. Real H II regions instead form within complex, clumpy 3D gas distributions with internal density gradients and are embedded in a multiphase ISM, as seen in high-resolution simulations of star cluster formation such as STARFORGE (M. Y. Grudić et al. 2021, 2022). Multicomponent photoionization models combining regions with different densities (e.g. Z. Martinez et al. 2025; B. Moreschini et al. 2026) have been shown to reproduce high-$z$ emission-line ratios more successfully than single-zone approximations by mitigating systematic biases introduced by uniform low-density assumptions. A physically motivated distribution of ionized gas densities may simultaneously reproduce observations in the N2-BPT and the S2-VO87 diagrams.

We further neglect turbulence, which can modify optical line ratios by changing the geometry and surface area of the H II region (Y. Jin, L. J. Kewley & R. Sutherland 2022a; Y. Jin, L. J. Kewley & R. S. Sutherland 2022b; L. Xing et al. 2026). In idealized turbulent H II region models, this effect increases low-ionization ratios such as N2 and S2, while decreasing R3 and O32. However, the predicted shifts are modest. In the most extreme case, L. Xing et al. (2026) find changes of $\sim 0.15$ dex in R3, N2, and S2, up to $\sim 0.4$ dex in O32 and $\sim 0.1$ dex in R23. These offsets are smaller than the shifts driven by the main physical mechanisms explored in this work, and turbulence is therefore expected to have a subdominant impact on our conclusions.

Lastly, our emission-line modelling is based solely on the contribution from discrete H II regions. However, as discussed in Section 4, the DIG can contribute to [S II] $\lambda\lambda$6717, 6731 and other low-ionization lines at $z \sim 0$ (R. L. Sanders et al. 2017; F. Mannucci et al. 2021; E. Congiu et al. 2023). As we focused on compact, high-sSFR systems locally and at high redshift, we do not expect our conclusions to be significantly changed by a potential DIG contribution.

### 6.2.4 *Caveats related to the star cluster modelling*

In resampling our star clusters, we adopt a cluster mass function with a power-law slope of $\alpha = -1.96$, as determined by N. Lahén et al. (2020). Our results show only a weak sensitivity to the precise value of the slope. Specifically, an order-of-magnitude change in cluster mass translates to a difference of only $\Delta \log U \approx 0.33$. Unless the resampled cluster masses shift by substantially more than an order of magnitude, any slope changes remain subdominant compared to other sources of H II region-to-H II region variation, such as age and gas-phase metallicity.

In one of the LUMEN versions, we explore the possibility of systematically more massive star clusters at high redshift. While observed star clusters at $z > 2$ appear generally more compact and massive than local counterparts (M. Dessauges-Zavadsky et al. 2017; A. Claeyssens et al. 2023; E. Vanzella et al. 2023; A. Adamo et al. 2024; M. Messa et al. 2024; L. Mowla et al. 2024; Abdurro'uf et al. 2025; A. Claeyssens et al. 2025, 2026; M. Messa et al. 2026), it is unclear if lower-mass clusters are underdetected rather than intrinsically absent, as observational constraints remain subject to resolution, sensitivity, and lensing selection effects. Thus it is not yet firmly established whether the underlying high-$z$ cluster population indeed favours higher masses. Nevertheless, theoretical work has shown that high gas surface densities, typical at high redshift, produce more massive star clusters (J. M. Kruijssen 2015; M. Reina-Campos & J. M. Kruijssen 2017; G. De Lucia et al. 2024b).

As observations (M. Dessauges-Zavadsky & A. Adamo 2018; A. Claeyssens et al. 2025, 2026) appear consistent with a local-like power law with $\alpha \approx -2$ at the high-mass end, we model this by increasing the minimum allowed cluster mass by 2–3 orders of magnitude from $10^3$ M$_\odot$ to $10^5$ and $10^6$ M$_\odot$. However, as shown by E. Vanzella et al. (2026), a flatter star-cluster mass function slope might help explain the excess of high-mass star clusters. As mentioned above, we do not expect our results to be particularly sensitive to the exact assumed shape of the cluster mass function, as long as there are significantly more star clusters exceeding masses of $10^5$–$10^6$ M$_\odot$ than at $z \sim 0$.

In LUMEN, the resampled star clusters are placed into the gas by assuming that the youngest clusters should reside in the densest gas. As a result, clusters are not connected self-consistently to the gas properties, as they would be if formed within their simulated birth clouds. Similarly, the ionization parameter is not computed directly from the gas densities predicted by the simulation, but is instead derived from the gas-phase metallicity using an assumed anticorrelation with the ionization parameter (D.

Carton et al. 2017). Considering that our methodology results in an ensemble of H II regions with diverse and realistic properties (see Section 4.1.1), we consider these reasonable approximations, but caution against analyses based on individual H II regions from the sample.

### 6.2.5 *Caveats related to dust effects*

Dust represents a further source of uncertainty in our modelling. In the ISM, the dust mass, grain-size distribution, composition, and spatial structure are strongly shaped by the ambient gas properties and environmental processing due to radiation fields, turbulence, and shocks. From their initial production in core-collapse supernovae and AGB stars, dust grains evolve through growth and processing in dense gas via accretion, coagulation, and chemical modification, and are subsequently destroyed by shocks, sputtering, shattering, and photo-processing (B. T. Draine 2003). These processes are not followed in TNG50. As mentioned in Section 3.3, we do not apply line-of-sight interstellar attenuation to the predicted emission-line strengths, since our analysis focuses on ratios of closely spaced emission lines for which differential dust attenuation is expected to be minimal. Dust internal to the H II regions is included through our CLOUDY models, although the exact modelling choices carry their own uncertainties. Fig. 13 shows the impact of choosing different settings for the dust-to-metal mass ratio and the grain-size distribution on the predictions in the N2-BPT diagram. We applied these changes to the model versions with high star-cluster masses, for which the impact is particularly strong.

In the top panel, we decreased the dust-to-metal mass ratio from $\xi_d = 0.3$ to 0.1, motivated by the potentially low dust content expected for high-$z$ galaxies, where grains have had limited time to grow and accumulate (A. K. Inoue 2003; R. S. Asano et al. 2013; A. Rémy-Ruyer et al. 2014; G. Popping, R. S. Somerville & M. Galametz 2017). Although both observational and theoretical studies generally find that $\xi_d$ decreases with decreasing metallicity and increasing redshift, the inferred scatter is large, spanning 2 dex from $\xi_d \sim 0.01$ to 1 (P. De Vis et al. 2019; Q. Li , D. Narayanan & R. Davé 2019). Lowering $\xi_d$ increases the cooling efficiency because a larger fraction of metal coolants remains in the gas phase. The resulting decrease in electron temperature reduces the rate of collisional excitation, which weakens both [N II]$\lambda$6584 and [O III]$\lambda$5007 emission and therefore lowers N2 and R3 (J. C. Shields & R. C. Kennicutt 1995; J. Gutkin et al. 2016). Since oxygen is refractory, it is significantly depleted onto dust grains with increasing $\xi_d$. At $\xi_d = 0.1$ the gas thus contains proportionally more oxygen than in our fiducial configuration with $\xi_d = 0.3$, which effectively decreases N/O. This higher O abundance partly offsets the reduced excitation for [O III]$\lambda$5007, resulting in a proportionally stronger drop in N2. Overall, due to the decline of both N2 and R3, we are not able to reproduce most observed galaxies at $z = 3$–8. The same simultaneous reduction in N2 and R3 appears across all LUMEN variants with $\xi_d = 0.1$, meaning that according to the assumptions of our modelling, a universally low $\xi_d$ is incompatible with the observations.

In the bottom panel, we replaced the Milky Way-like ISM grain-size distribution with an Orion-like distribution, which features fewer small grains (J. C. Weingartner & B. T. Draine 2001). Ionizing photons are absorbed efficiently by small grains, so reducing their abundance increases the strengths of H $\alpha$ and H $\beta$. Because

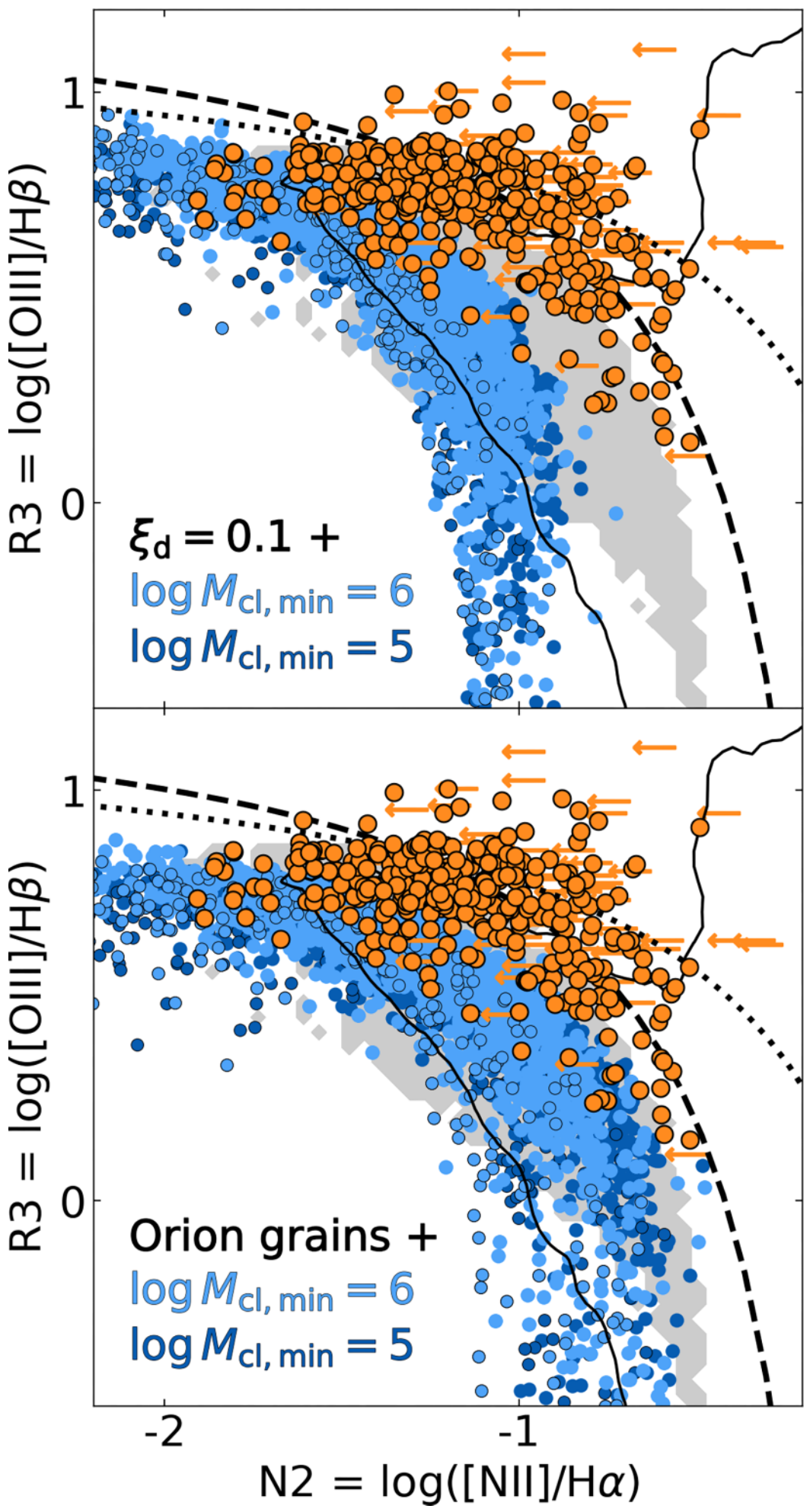

**Figure 13.** Same as Fig. 7, comparing versions of LUMEN between $z = 3$–8 with increased star-cluster masses and a $\xi_d$ of 0.1 (top) and an Orion-like dust-grain size distribution (bottom).

[O III]$\lambda$5007 is collisionally excited and more sensitive to changes in temperature, the altered grain distribution also modifies the heating-cooling balance. Together, these effects eliminate the increase in R3 seen when using the CLOUDY ISM-type grains in the massive star cluster models. While one might argue that a star-forming region like Orion provides a better dust approximation when modelling H II regions, extinction curves of star-forming regions vary substantially between the Milky Way and the Large and Small Magellanic Clouds, with some even showing an excess of small dust grains (K. D. Gordon, D. Calzetti & A. N. Witt 1997; K. D. Gordon et al. 2003; G. De Marchi & N. Panagia 2014). Regions dominated by strong radiation fields or shocks typically show steeper far-UV slopes due to an enhanced fraction of small grains, while environments where grain growth or coagulation dominates exhibit flatter curves with larger grains. Low metallicities in early systems potentially favour a grain population skewed towards even larger sizes from stellar ejecta rather than ISM grain growth, yet their intense radiation fields and feedback processes promote grain destruction. Given these competing effects, no single grain-size distribution is universally representative for all H II regions in all galaxies at all redshifts. Dust-evolution simulations support this picture. S. Aoyama et al. (2017) and K. C. Hou et al. (2019) show that the small-to-large grain abundance ratio

depends on time, metallicity, redshift, and ISM phase, reflecting the changing balance between shattering, accretion, and coagulation. A fixed Milky Way-like or Orion-like grain-size distribution is therefore unlikely to capture the full range of dust conditions in high-$z$ H II regions. Instead, real galaxies likely host a range of dust grain size distributions shaped by their specific radiation and feedback environments.

### 6.3 Comparison to previous theoretical high-$z$ emission-line studies

In Section 5, we compared predictions from our LUMEN emission-line catalogues for $z = 3$–8 to recent *JWST* observations in the N2-BPT, S2-VO87, and O32–R23 diagrams. We found that reproducing the full distribution of observed line ratios requires a combination of harder ionizing radiation, higher ionization parameters, and elevated N/O abundances. A number of previous theoretical studies have explored the physical origin of the N2-BPT and O32–R23 offsets, using approaches ranging from semi-analytic modelling to large-volume simulations and direct photoionization calculations. Below, we summarize their main conclusions and relate them to our findings.

#### *6.3.1 Comparison to modelling from M. Hirschmann et al. (2023b)*

Our modelling builds on the framework of M. Hirschmann et al. (2017, 2019), M. Hirschmann et al. (2023a), and M. Hirschmann et al. (2023b), in which integrated galaxy emission was represented by an ensemble of H II regions sharing the same galaxy-wide properties. In that approach, the effective ionizing photon rate $Q$ was mapped directly to the global SFR, so that higher-SFR galaxies contained regions with larger $Q$ and higher ionization parameters. This implicitly assumed that galaxies with higher SFRs form the same number of H II regions as lower-SFR galaxies, but with each region powered by a larger $Q$ and therefore a higher ionization parameter. However, this assumption does not hold if, in real galaxies, a higher global SFR instead produces a larger number of H II regions without changing their $\log U$. While the resulting integrated emission would appear brighter, the galaxy spectra would not exhibit higher-excitation line ratios, since multiple low-ionization regions cannot reproduce the emission of one highly ionized nebula. The ionization state therefore depends not only on the total SFR, but also on how newly formed stellar mass is distributed among star clusters.

LUMEN addresses this limitation by explicitly sampling the star clusters from a realistic mass distribution, allowing $\log U$ to respond to changes in the cluster mass function. The framework further captures spatial variations in metallicity and ionization conditions, which result in H II regions with diverse properties. Our LUMEN predictions are partially consistent with M. Hirschmann et al. (2023b). In post-processing the TNG50, ILLUSTRISTNG100, and ILLUSTRISTNG300 simulations with photoionization models, they found that elevated ionization parameters, linked to higher specific SFRs and increased global gas densities in early galaxies, may drive a continued increase in R3 out to $z = 7$. In this work, we have shown that the evolution of line ratios is instead driven by multiple processes, including an increase in $\log U$ driven by high star-cluster masses and potentially elevated ionized gas densities. See Appendix B for a detailed comparison.

#### *6.3.2 Comparison to other works*

Several other recent studies have investigated high-$z$ optical line ratios using simulation frameworks. B. E. Backhaus et al. (2024) compared the M. Hirschmann et al. (2023b) catalogues to an early CEERS sample of 27 galaxies at $z > 5$, but found observed R3 values (0.3–1) generally lower than predicted (0.7–1.3), suggesting that the simulated ionization parameters might be too high. S. Fujimoto et al. (2023) compared CEERS galaxies at $z = 7$–8 to the ILLUSTRISTNG-based predictions from M. Hirschmann et al. (2023b), an adapted method applied to the SANTA CRUZ SAM, and predictions from the FLARES simulations (S. M. Wilkins et al. 2023). They concluded that ILLUSTRISTNG and the SANTA CRUZ SAM reproduce the observed R3 values, whereas FLARES does not predict R3 values above 0.4.

H. Katz et al. (2023) used the SPHINX radiation-hydrodynamic simulations combined with CLOUDY modelling to predict emission lines for $z = 4.6$–10 galaxies. Their results showed a redshift evolution of the N2-BPT sequence driven mainly by decreasing metallicity. The effect of decreasing metallicities is also captured in our approach, but we find it to be insufficient to explain any N2-BPT evolution. More in line with our findings, they also note that galaxies matching NIRSpec samples in the N2-BPT and O32–R23 diagrams at $z > 5$ exhibit particularly high ionization parameters due to extremely dense gas. S. Kusmic et al. (2026) post-processed TECHNICOLORDAWN galaxies with CLOUDY and found little evolution in O32 or R3 at $z > 5$, with R3 systematically lower than observed. They attributed this discrepancy in part to overly high oxygen abundances and suggested that higher ionization parameters are required to match *JWST* data.

The MEGATRON RHD simulations (H. Katz et al. 2025; N. Choustikov et al. 2026) model H II regions on the fly for galaxies at $z > 8.5$, if the Jeans length is resolved. In unresolved dense regions, cell emission is instead replaced with CLOUDY models. While the resulting galaxies overlap with observed galaxies in the O32–R23 diagram, as shown in N. Choustikov et al. (2026), they struggle to reproduce galaxies with simultaneously high R23 (0.5–1) and low O32 ($< 1$). They argue that these objects are likely post-starburst galaxies with higher metallicities, which are not contained in the MEGATRON simulation volume. In LUMEN, these R23–O32 combinations arise in metal-rich systems ($\log Z/Z_{\odot} > 0.5$) when either combined stellar $\alpha$-enhancement and a higher $m_{\rm up}$ or increased star-cluster masses are included. N. Choustikov et al. (2026) note that above O32 values of 1.25, O32 is strongly correlated with gas densities, largely due to the low critical density of the [O II] line. As a result, their most extreme O32 values are produced by the highest gas densities, similar to our findings. Extreme R3 values (R3 $> 1$) in the MEGATRON suite appear only under a variable IMF, although N. Choustikov et al. (2026) do not further explore the physical origin of this behaviour.

Finally, W. McClymont et al. (2025) applied the COLT Monte Carlo radiative transfer solver to the THESAN-ZOOM simulations. Their predictions show O32 and R3 values broadly consistent with galaxies from JADES, but they do not reach the more extreme observed ratios and suffer from limited statistics at stellar masses above $10^8$ $M_{\odot}$.

Overall, no existing simulation-based model has yet been shown to reproduce the full set of $z > 3$ optical line-ratio constraints considered here. Previous studies can match selected parts of the observed parameter space, such as elevated R3 values, regions of the O32–R23 plane, or aspects of the high-$z$ N2-BPT offset, but none simultaneously explains the full combination of high R3, high O32, and the observed N2-BPT behaviour across the relevant redshift and stellar-mass range. As for prior work at $z \sim 2$, commonly invoked drivers include elevated ionization parameters, higher gas densities, and lower metallicities. No study at $z > 3$ has explicitly modelled elevated N/O or hardened ionizing spectra from an increased IMF upper-mass cutoff or $\alpha$-enhanced populations. Observational interpretations also differ, with some attributing BPT evolution to stellar $\alpha$-enhancement (R. L. Sanders et al. 2023a; A. E. Shapley et al. 2025) and others to high ionization parameters and low metallicities (A. J. Cameron et al. 2023). This diversity of interpretations highlights the need for a framework like LUMEN, which enables the first comprehensive assessment of all proposed evolutionary drivers within a cosmological context.

## 7 CONCLUSION

In this paper, we introduced LUMEN, a new methodology to model spatially distributed H II regions in cosmological simulations. The approach preserves the cosmological context from low to high redshift, captures the internal diversity of nebular conditions within galaxies, and links each star cluster to the local gas environment to model H II region emission lines. By sampling individual star clusters with realistic masses and assigning to each a stellar population and photoionization model, LUMEN bridges cosmological and ISM scales in a statistically robust way. The resulting framework provides a controlled environment to test physical mechanisms proposed to explain elevated high-$z$ emission-line ratios recently reported by *JWST*. Before extending our models to high redshift, we validated the LUMEN Base model extensively at $z = 0$. At the level of individual H II regions, LUMEN matches observed distributions of properties such as ionization parameter, size, and H $\alpha$ luminosity. Galaxy-integrated spectra further reproduce local optical line-ratio sequences and strong-line tracers for star formation and metallicity. This confirms that LUMEN produces physically credible nebular emission across multiple scales.

We then explored model variations targeting processes commonly invoked at high redshift, including harder spectra (via binary stars, higher IMF upper-mass cutoff, stellar $\alpha$-enhancement, and AGNs), elevated ionization parameters (via higher cluster masses and high ionized gas densities), enhanced N/O ratios, and LyC leakage. This enabled a systematic assessment of how these mechanisms affect R3, N2, S2, and O32 in the N2-BPT, S2-VO87, and O32–R23 diagrams at $z = 3$–8. The main results are:

(i) Even with relatively extreme stellar $\alpha$-enhancement, adopting $Z_\star = Z_{\rm gas}/5$, spectral hardening alone cannot reproduce the bulk of the observations. Both $\alpha$-enhanced stellar populations and a higher IMF upper-mass cutoff harden the ionizing spectrum and mildly increase N2, S2, R3, and O32. Together they reproduce moderately elevated N2, S2, and R3 values, but still fail to reach the most extreme line ratios observed at $z = 3$–8. Including binary-star SEDs from BPASS has little additional impact.

(ii) While weak AGNs have little effect, strong AGN contributions move a significant number of predicted $z = 3$–4 line ratios into the observed regime. However, unless TNG50 underpredicts the abundance of high-$z$ AGNs, AGN contributions cannot explain the extreme line ratios at $z > 4$. Furthermore, most strong AGNs would likely be identified and excluded from observational samples.

(iii) Reproducing the most extreme R3 and O32 requires elevated ionization parameters. These can be achieved by increasing the lower-mass cutoff of the star-cluster mass function to $10^5$–$10^6$ $M_\odot$, potentially consistent with *JWST* detections.

(iv) Increasing the ionized gas density from $10^2$ to $10^3$–$10^4$ cm$^{-3}$ also increases the ionization parameter and enhances R3, O32, and N2. The $10^4$ cm$^{-3}$ model matches N2-BPT and O32–R23 observations but suppresses S2, which creates tension with S2-VO87 constraints. Multicomponent density models may alleviate this.

(v) Matching the highest N2 values without reducing S2 may require enhanced N/O, corresponding to $\log(\rm N/O) \sim -0.5$ to 0.5 at $12 + \log(\rm O/H) \geq 8$, consistent with observed high-$z$ nitrogen emitters.

(vi) LyC leakage from density-bounded regions produces only modest O32 increases at $\log U < -2$ and cannot explain high-$z$ observations without elevated ionization parameters.

(vii) Overall, no single mechanism can explain the full range of high-$z$ emission-line ratios by itself. To test whether a physical picture combining multiple processes can reproduce all observed trends, we constructed one additional 'Master' model. This model includes harder ionizing radiation due to an IMF upper-mass cutoff of 300 $M_\odot$ and $\alpha$-enhanced stellar populations, increased ionization parameters driven by massive star clusters, and elevated N/O abundances. The 'Master' model successfully reproduces the most extreme R3, N2, S2, and O32 ratios at $z \geq 3$ collectively across the three diagnostic diagrams, indicating that the observed galaxies are likely subject to varying combinations of these processes. Based on this model, we provide new demarcation lines for star-forming galaxies in the N2-BPT and the S2-VO87 diagrams.

*JWST* will continue to expand the sample of early line-emitting galaxies and provide increasingly detailed constraints on the physical conditions of the high-$z$ ISM. Flexible theoretical frameworks such as LUMEN will be essential for interpreting these data, as they offer testable predictions for a range of physical mechanisms. Because LUMEN reproduces both integrated galaxy spectra and the underlying H II-region statistics, it enables a broad range of applications. Forthcoming studies will use LUMEN-generated emission-line data for TNG50 and COLIBRE galaxies (J. Schaye et al. 2026) to test the reliability of direct-temperature diagnostics under different emission-line modelling assumptions and to quantify biases in metallicity gradients derived from strong-line diagnostics. Future work will also adapt LUMEN for application to large-volume, lower-resolution simulations and semi-analytic models, enabling the exploration of volumes comparable to those probed by cosmological surveys. More broadly, this type of modelling will be essential for interpreting not only *JWST* spectra, but also observations from forthcoming spectroscopic facilities such as the ELT and instruments such as MOSAIC.

## ACKNOWLEDGEMENTS

LS, MH, AP, and MF acknowledge funding from the Swiss National Science Foundation (SNSF) through PRIMA grant PR00P2 193577, 'From cosmic dawn to high noon: the role of black holes for young galaxies'. The Flatiron Institute is supported by the Simons Foundation. ECL acknowledges support from a Science and Technology Facilities Council (STFC) Webb Fellowship (ST/W001438/1). AF acknowledges support from the Istituto Nazionale di Astrofisica (INAF) through the Ricerca Fondamen-tale INAF 2024 MiniGrant RSN1 (grant 1.05.24.07.01), 'The pc-scale view of H II regions in M33', and the INAF project 'Very Large Telescope – Multi-Object Optical and Near-infrared Spectrograph (VLT-MOONS)' (CRAM 1.05.03.07). NL was supported by a Gliese Fellowship at the Zentrum für Astronomie, Universität Heidelberg, Germany, and acknowledges the computational resources provided by the Max Planck Computing and Data Facility in Germany.

## DATA AVAILABILITY

The data underlying this article will be shared upon reasonable request to the corresponding author. The LUMEN model has primarily been developed by LS, MH, and AP. We kindly ask that any inquiries regarding the use of LUMEN or requests for collaboration be directed to these authors. The ILLUSTRISTNG simulations are publicly available and accessible at www.tng-project.org/data (D. Nelson et al. 2019a).

## APPENDIX A: LINE-RATIO DIAGRAMS AT $z = 2$

Here we provide a comparison of the LUMEN model suite to observations at $z \sim 2$. Fig. A1 shows the predicted locations of simulated galaxies at $z = 2$ in the classical N2-BPT diagram. The model configuration is identical to that used in Fig. 7. In addition, we include a pre-*JWST* empirical fit at $z \sim 2.3$ with a $1\sigma$ scatter (C. C. Steidel et al. 2014; dashed orange line with shaded yellow area), as well as observations and [N II]$\lambda$6584 upper limits at $z = 1.4$–2.7 from the AURORA, CEERS, and JADES programs. The *JWST* observations are broadly within or below the C. C. Steidel et al. (2014) relation and its scatter, except for three extreme R3 values above 1 and one N2 value at $\sim -0.5$ and R3 $\sim 0.5$.

Similar to $z = 3$–8, the Base and BPASS models underpredict the observed distribution. Variants with a higher IMF upper-mass cutoff or stellar $\alpha$-enhancement shift the population to slightly higher R3 and N2, reproducing the high-R3, low-N2 locus. However, they remain below the most extreme R3 values and do not reach the high-N2 tail extending into the composite region between the AGN and starburst classification curves. Combining both mechanisms shifts the locus above the C. C. Steidel et al. (2014) relation, with some models reaching the most extreme observed R3 values and slightly improving the match to the high-N2 tail. Both weak and strong AGNs can produce R3 and N2 values comparable to or higher than those observed at $z = 2$, but the observed R3 $> 1$ systems occur at lower N2 than predicted. Hidden AGNs may elevate R3 and N2 in individual systems, but are unlikely to explain the full $z \sim 2$ population. Increasing the minimum star-cluster mass reproduces the high-R3 tail of C. C. Steidel et al. (2014), reaching R3 $\sim 1$, but lowers N2 with increasing R3, failing to match the high-N2, low-R3 regime and the most extreme R3 values. Increasing the ionized gas density shifts the population to higher R3 and N2. $n_{\rm H} = 10^3\,{\rm cm}^{-3}$ reproduces the C. C. Steidel et al. (2014) locus, while $10^4\,{\rm cm}^{-3}$ reaches the most extreme values but is likely unrealistic for typical $z \sim 2$ systems. Nitrogen enhancement similarly raises N2, though the mild model is insufficient and the strong model overshoots, suggesting an intermediate enhancement. Overall, no single mechanism reproduces the full distribution in the N2-BPT diagram at $z \sim 2$. Because the observed line ratios are less extreme than at $z = 3$–8, the 'Master' model would systematically overpredict the distribution. Instead, the data would likely be consistent with a range of less extreme models combining partially degenerate effects from harder ionizing radiation, AGNs, increased star-cluster masses, higher ionized-gas densities, and elevated nitrogen abundances.

Similar trends are seen in the S2-VO87 diagram at $z = 2$ (Fig. A2), which compares LUMEN predictions to observations at $z = 1.4$–2.7. At higher redshift, the Base and BPASS models fail to reproduce galaxies above the L. J. Kewley et al. (2001b) line. Harder ionizing spectra from a higher IMF upper-mass cutoff or stellar $\alpha$-enhancement match most R3 and S2 values, missing only the most extreme R3 cases, while their combination shifts the locus closer to these points. Both weak and strong AGNs reproduce the overall distribution, though it is unlikely that AGN contamination dominates the full sample. Increasing the minimum cluster mass raises R3 at fixed or slightly lower S2, matching systems just above the classification line but not the most extreme R3 values. Higher gas densities increase R3 without raising S2 at $n_{\rm H} = 10^3\,{\rm cm}^{-3}$ and suppress S2 at $10^4\,{\rm cm}^{-3}$ due to exceeding the [S II] critical density. Enhanced N/O similarly reduces S2 through abundance rescaling at fixed metallicity. As before, the observations are most consistent with combinations of partially degenerate mechanisms rather than a single dominant driver.

Fig. A3 shows the O32–R23 diagram at $z = 2$, compared to observations at $z = 1.4$–2.7. Most data are reproduced by the Base and BPASS models, except for systems with O32 $\gtrsim 0.9$ extending to $\sim 1.5$ and R23 $\sim 1$. A higher $m_{\rm up}$ matches high-O32 values up to $\sim 1.1$, while stellar $\alpha$-enhancement yields a smaller increase. Combining both improves agreement across O32 and R23. AGNs may account for a few individual high-O32 systems. Models with higher cluster masses or $n_{\rm H} = 10^4\,{\rm cm}^{-3}$ overshoot the sequence, whereas $n_{\rm H} = 10^3\,{\rm cm}^{-3}$ is more consistent. LyC leakage provides only a modest O32 boost. Overall, multiple mechanisms can reproduce the observed $z = 2$ O32–R23 distribution, and current data do not strongly constrain the dominant driver.

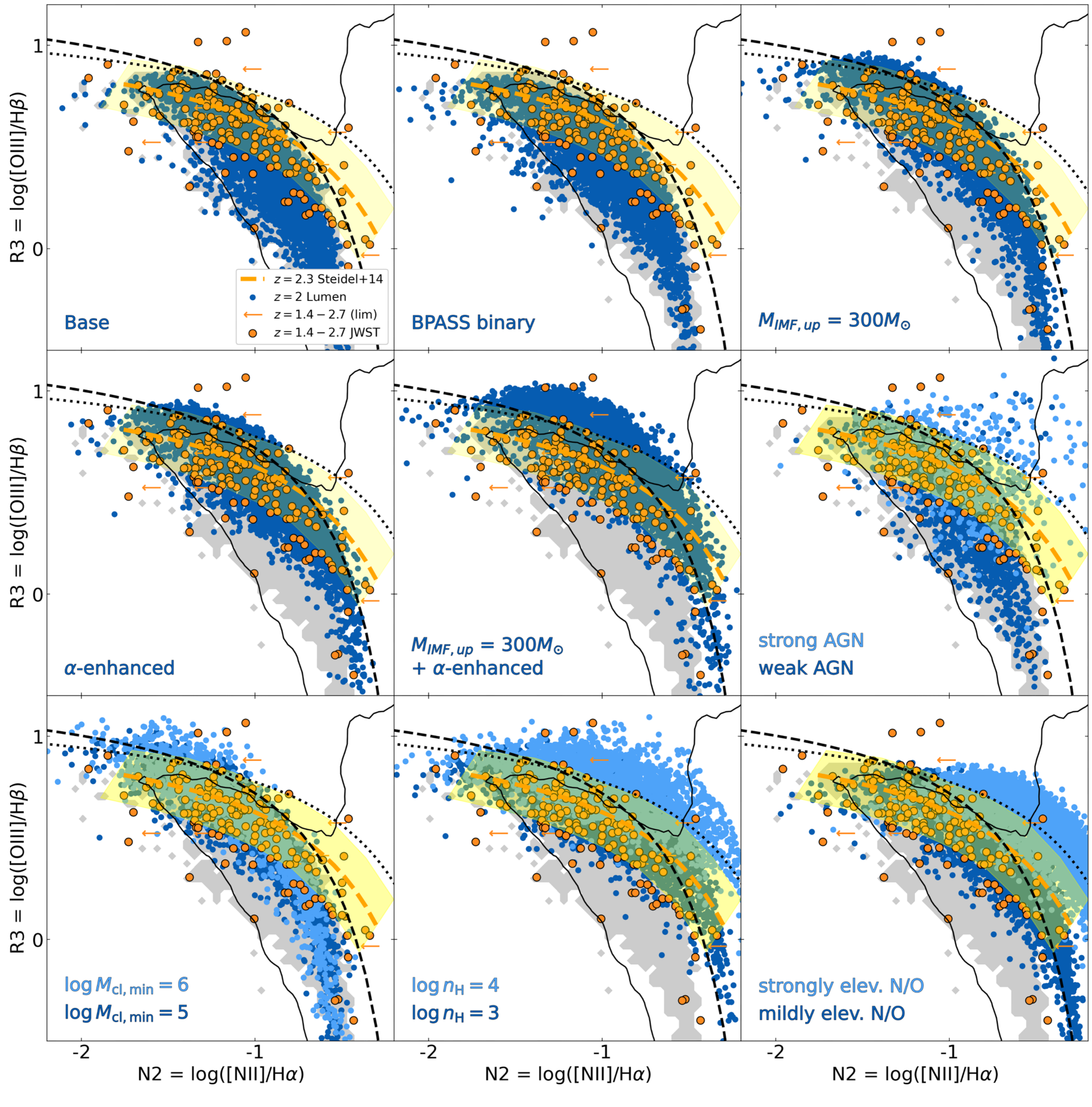

**Figure A1.** As in Fig. 7, but at $z = 2$. The observational comparison includes SDSS galaxies (black outline), local ionizing source criteria (dashed and dotted lines), the empirical fit at $z \sim 2.3$, with a $1\sigma$ scatter (C. C. Steidel et al. 2014; dashed orange line with shaded yellow area), as well as *JWST* observations and [N II]λ6584 upper limits at $z = 1.4$–$2.7$ (orange).

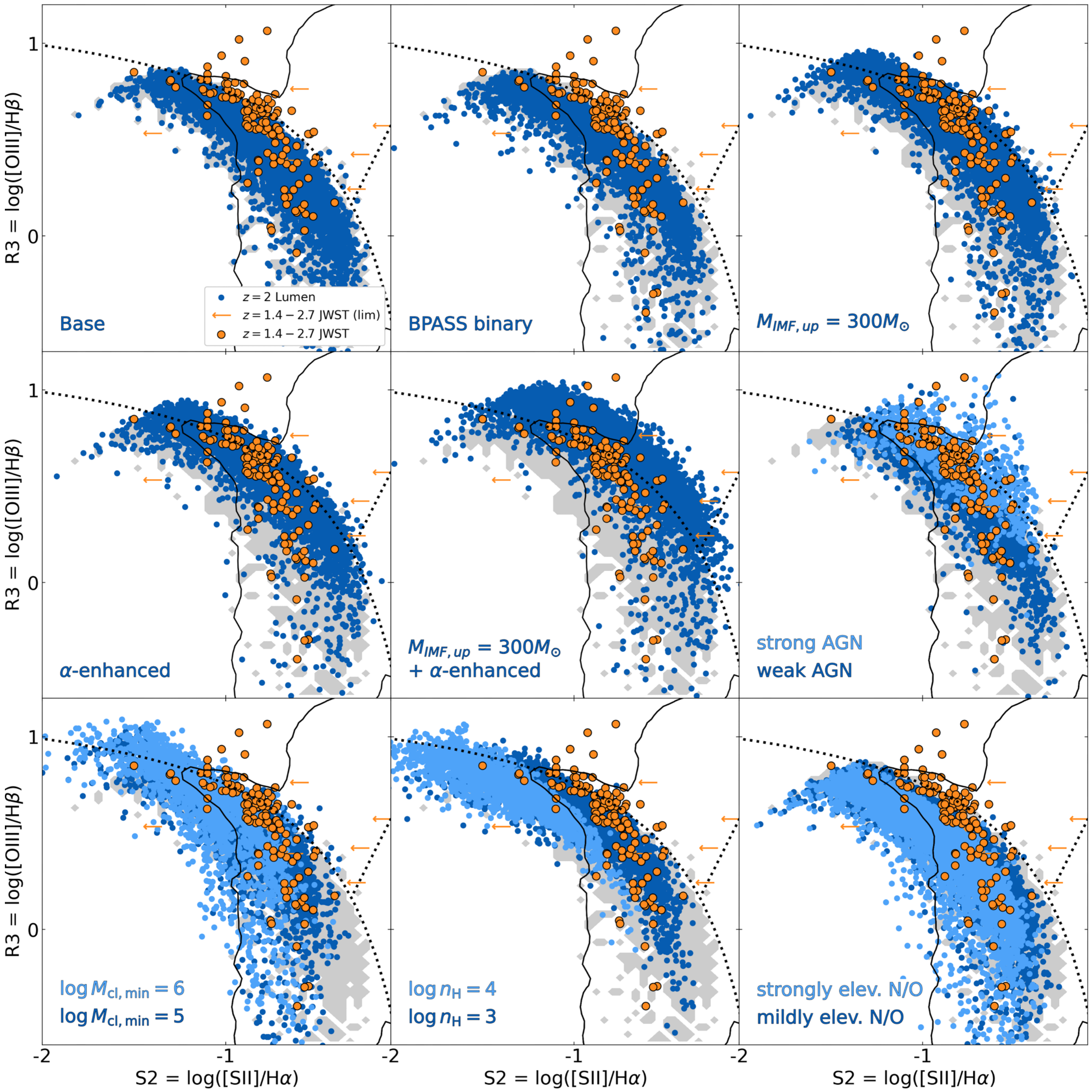

**Figure A2.** As in Fig. 8, but at $z = 2$. Included for comparison are SDSS galaxies (black outline), local ionizing source criteria (dashed and dotted lines), as well as *JWST* observations at $z = 1.4$–$2.7$ (orange).

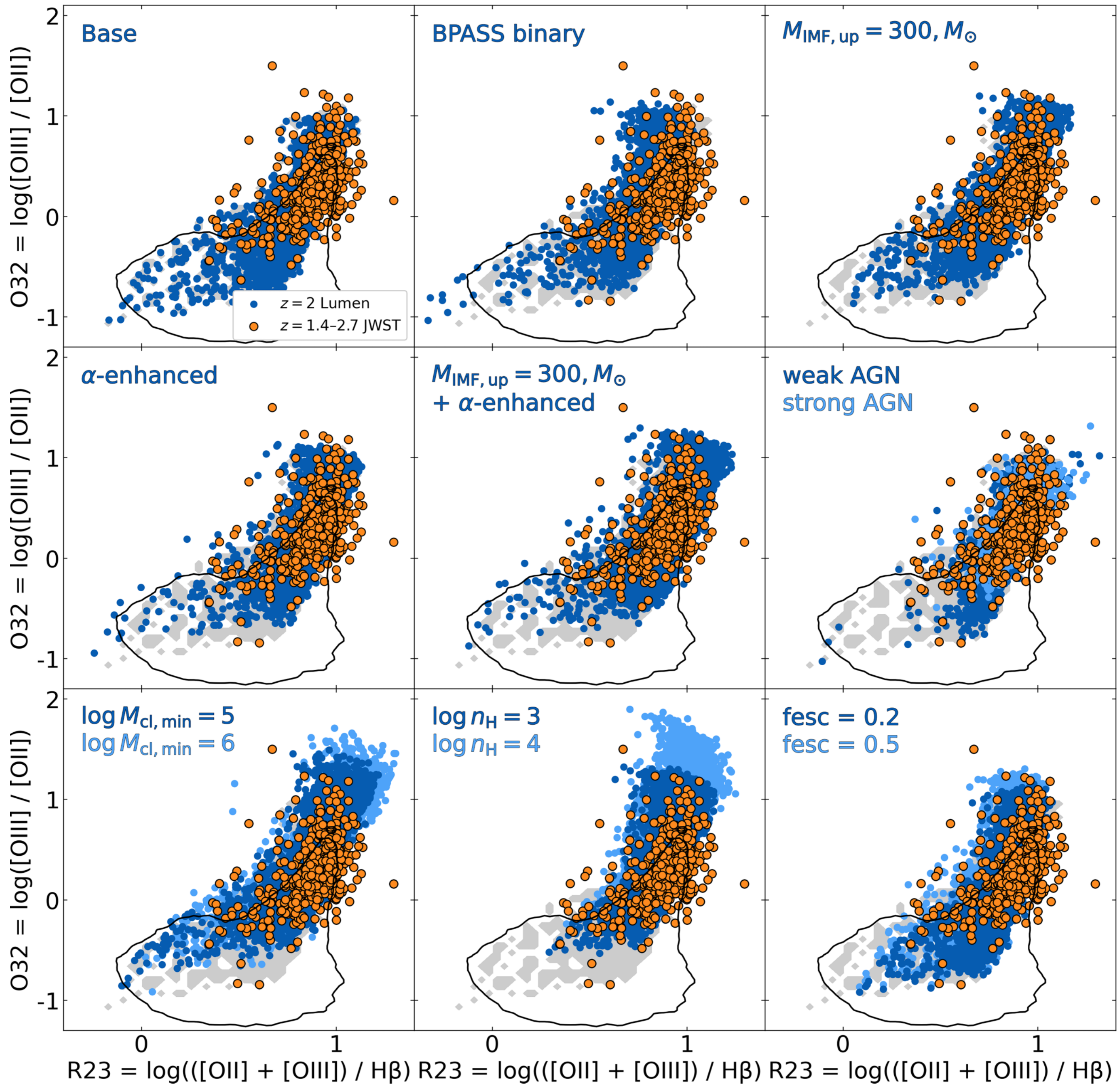

**Figure A3.** As in Fig. 9, but at $z = 2$ with *JWST* observations between $z = 1.4$and 2.7.

## APPENDIX B: COMPARISON WITH HIRSCHMANN ET AL. (2023)

In this section, we compare our results to the predictions obtained for TNG50 using the M. Hirschmann et al. (2023b) method for the H II component. Similar to Fig. 5, Fig. B1 shows the N2-BPT, S2-VO87, [O I]λ6300/H α-VO87, and O32–R23 diagrams at $z = 0$, here containing the predictions for galaxies with $M_\star > 10^8\,{\rm M_\odot}$ from M. Hirschmann et al. (2023b; pink) in addition to LUMEN. Because only one photoionization model is assigned to each galaxy, the resulting predictions follow the discrete grid of the J. Gutkin et al. (2016) photoionization models. Consequently, this approach cannot reproduce the same diversity of emission-line predictions as LUMEN. Overall, the two approaches cover similar regions in the diagrams, but the models of M. Hirschmann et al. (2023b) extend further into regions associated with lower metallicity, i.e. low N2, low S2, low [O I]λ6300/H α-VO87, and low R23. This is likely because the photoionization models are coupled using the global interstellar metallicity of each galaxy, which is typically lower than the star-forming-gas metallicity adopted by LUMEN.

Fig. B2 shows the H α and [O III] line luminosity functions, similar to Fig. 11. The M. Hirschmann et al. (2023b) method underpredicts the observed H α function from A. Covelo-Paz et al. (2025) by around 1 dex and generally predicts lower luminosities than LUMEN. This difference can be partially attributed to the high ionization parameters assigned to galaxies by the M. Hirschmann et al. (2023b) method, typically above $\log U =$

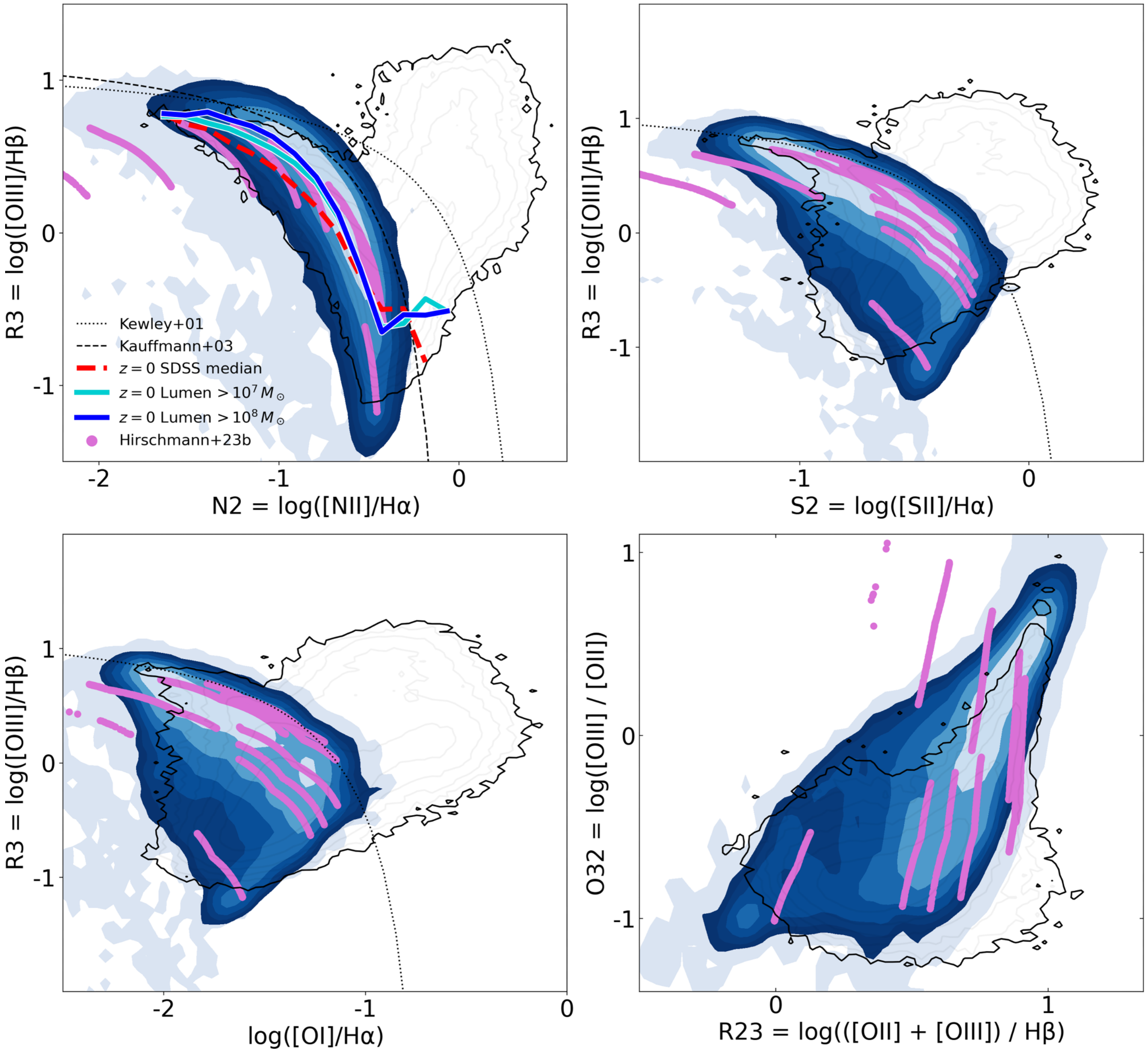

**Figure B1.** As in Fig. 5, but including predictions from M. Hirschmann et al. (2023b) for TNG50 (pink).

−2, which are associated with greater dust attenuation in the J. Gutkin et al. (2016) models. The effect is similar, but less severe for [O III], as [O III] luminosities increase due to the greater availability of $O^{2+}$ in the gas and the fact that its collisional excitation depends on the gas temperature. In Fig. B3, we compare the 'Master' model and M. Hirschmann et al. (2023b) predictions at $z = 3$–8 in the N2-BPT, S2-VO87, and O32–R23 diagrams. While M. Hirschmann et al. (2023b) can reproduce high R3 values, it overpredicts R3 while underpredicting N2 and S2. Both models reach the most extreme O32 values but miss most of the lower O32 values. Together, these plots show again that while elevated ionization parameters likely play a part in the evolution of emission-line signatures, they are not sufficient to explain all line-ratio signatures at high redshift. This underscores the importance of the improvements provided by LUMEN, which can explore many different physical processes and reproduce diverse line signatures.

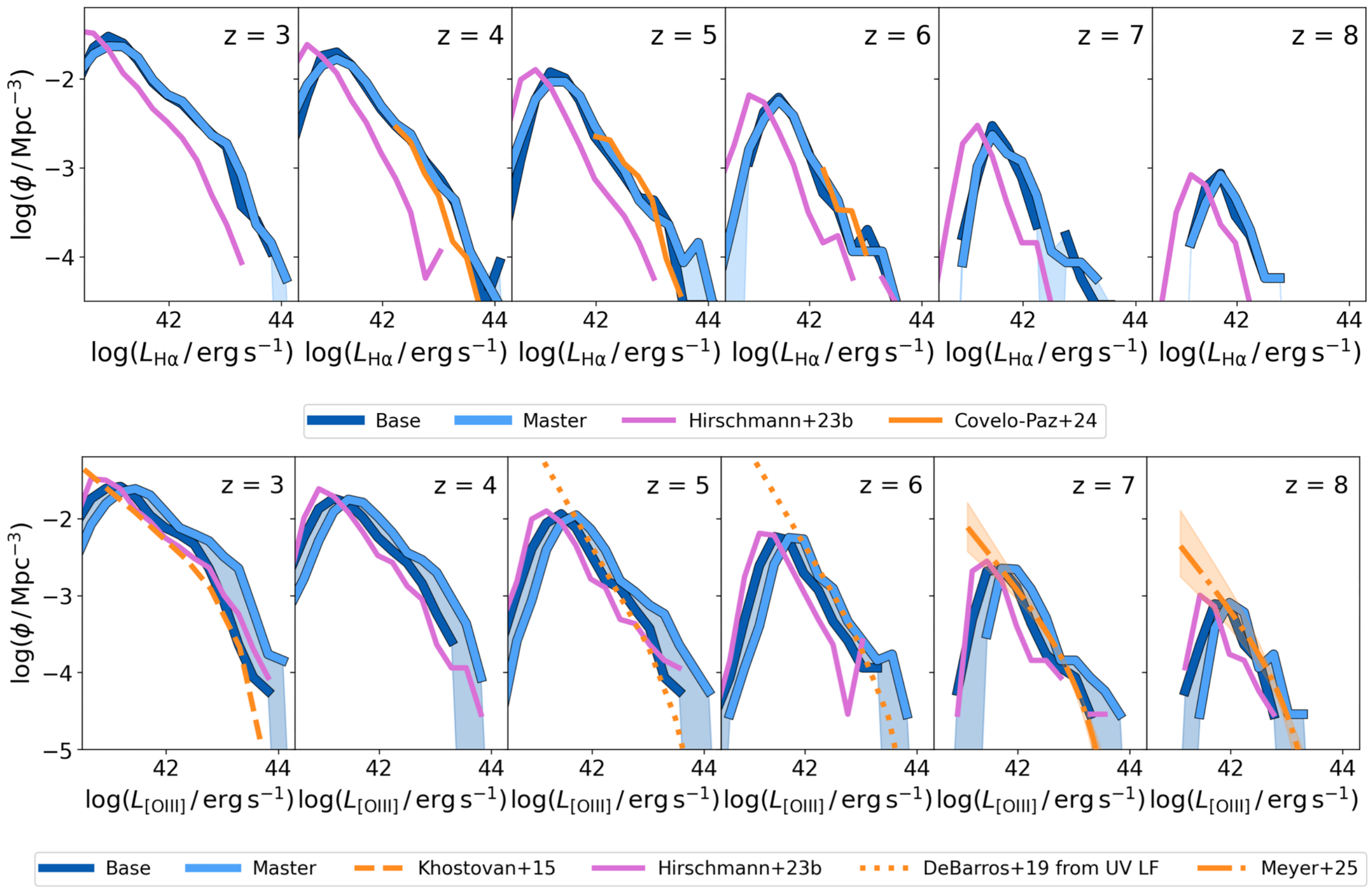

**Figure B2.** As in Fig. 11, but containing the prediction from M. Hirschmann et al. (2023b) for TNG50 (pink).

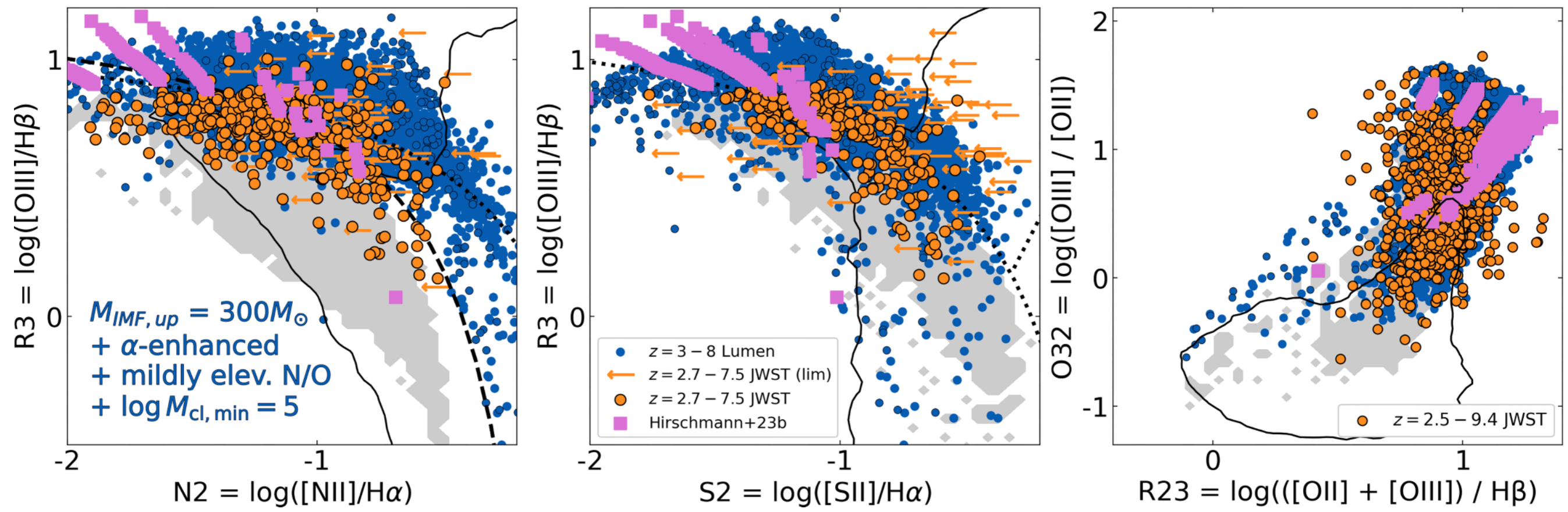

**Figure B3.** As in Fig. 10, but including predictions from M. Hirschmann et al. (2023b) for TNG50 (pink).

This paper has been typeset from a TEX/LATEX file prepared by the author.